\documentclass[a4paper,11pt]{article}
\usepackage{jheppub} % for details on the use of the package, please see the JINST-author-manual
\usepackage{amsmath}
\usepackage{booktabs}
\usepackage{colortbl}
\usepackage{siunitx}
\usepackage{subcaption}
\usepackage{slashed}
\usepackage{xspace}
\usepackage{placeins}

\usepackage[breakable]{tcolorbox}
\tcbuselibrary{minted}
\tcbuselibrary{skins}

\newcommand{\gosam}{\textsc{GoSam}\xspace}
\newcommand{\sherpa}{S\protect\scalebox{0.8}{HERPA}\xspace}

\newcommand{\amegic}{A\protect\scalebox{0.8}{MEGIC}\xspace}
\newcommand{\comix}{C\protect\scalebox{0.8}{OMIX}\xspace}
\newcommand{\CSS}{CSS\protect\scalebox{0.8}{HOWER}\xspace}
\newcommand{\alaric}{A\protect\scalebox{0.8}{LARIC}\xspace}

\newcommand{\rivet}{R\protect\scalebox{0.8}{IVET}\xspace}
\newcommand{\hepmc}{\textsc{HepMC}\xspace}

\newcommand{\MEPSatNLO}{M\protect\scalebox{0.8}{E}P\scalebox{0.8}{S}@N\protect\scalebox{0.8}{LO}\xspace}
\newcommand{\MEPSatLO}{M\protect\scalebox{0.8}{E}P\scalebox{0.8}{S}\xspace}

\DeclareMathOperator{\tr}{tr}

\newtcblisting{yamllst}[1]{colframe=gray!40, colback=gray!10, listing only, minted language=yaml, minted options = {fontsize=\footnotesize}, lower separated = false, comment style={size=fbox}, after={\par\noindent},#1}

\title{\boldmath Triple Higgs Production in Higgs Effective Field Theory with Parton Shower Effects}
\author[a]{Jens Braun,}
\author[a]{Gudrun Heinrich,}
\author[b]{Daniel Reichelt}

\affiliation[a]{Institute for Theoretical Physics, Karlsruhe Institute for Technology (KIT), Wolfgang-Gaede-Str.~1, 76131 Karlsruhe, Germany}
\affiliation[b]{CERN, Theoretical Physics Department, CH-1211 Geneva 23, Switzerland}

\emailAdd{j.braun@kit.edu}
\emailAdd{gudrun.heinrich@kit.edu}
\emailAdd{d.reichelt@cern.ch}

\preprint{{\small KA-TP-35-2026, P3H-26-096, CERN-TH-2026-312}}


\abstract{We present predictions for $gg\to HHH$ within Higgs Effective Field Theory (HEFT), including the full set of leading operators in the HEFT power counting. The calculation is based on a new interface between the one-loop provider \textsc{GoSam} and the Monte Carlo program \textsc{Sherpa}, enabling automated calculations within EFT frameworks. We study the interplay between the different operators as well as parton shower effects.}

\begin{document}
\maketitle
\flushbottom

\section{Introduction}
 The Higgs potential is still the least explored part of the Standard
Model~(SM). While the Higgs boson mass is measured to per-mille accuracy,
the self-couplings of the Higgs boson --- in the SM completely fixed once the
Higgs mass and the vacuum expectation value $v$ are known --- are still
awaiting a more precise experimental confirmation.  Rapid progress has been
made since the onset of the LHC to constrain the trilinear Higgs coupling,
and the prospects for the high-luminosity (HL)-LHC are excellent, with a
discovery of Higgs boson pair production within reach.

Direct access to the quartic Higgs self-coupling $\lambda_4$ through the
measurement of triple Higgs production at the HL-LHC seems out of reach if
SM couplings are assumed. However, searches for this process have already
been performed~\cite{ATLAS:2024xcs,CMS:2026xkc}, and to derive reliable constraints on the quartic
coupling modifier $\kappa_4$ it is indispensable to have predictions for
this process within a consistent effective field theory~(EFT) framework.
In addition, the quartic coupling enters the NLO electroweak corrections to
Higgs boson pair production, which can be used to obtain further constraints
on $\kappa_4$.

The dominant production mechanism for triple Higgs production at the LHC is
gluon fusion, $gg \to HHH$, which proceeds via loops of massive quarks.
The first leading-order calculations of this process were performed in
Refs.~\cite{Plehn:2005nk,Binoth:2006ym}.
The $gg \to HHH$ amplitude in analytic form was recently provided
in Ref.~\cite{Campbell:2025llb}.

Next-to-leading-order~(NLO) QCD corrections were
studied in Ref.~\cite{Maltoni:2014eza} in an approximation called FT$_{\rm approx}$, which is based on the heavy top limit
rescaled with the exact leading-order amplitude for the virtual corrections, and with top-mass effects incorporated in the real corrections. This led to a K-factor of about 1.6 at $\sqrt{s}=14$\,TeV. The differential K-factor in NLO FT$_{\rm approx}$ is relatively flat for the triple Higgs invariant mass distribution up to about 2\,TeV.
NNLO QCD corrections have been calculated in
Refs.~\cite{deFlorian:2016sit,deFlorian:2019app}, starting from the heavy-top limit and reweighting the NNLO cross-section in the soft-virtual approximation by the exact LO result. Partial finite top mass eﬀects have been included by taking into account the NLO FT$_{\rm approx}$ results of Ref.~\cite{Maltoni:2014eza}, using two alternative reweighting procedures, finding an overall NNLO K-factor of about 1.7 at total cross section level and a reduction of the scale uncertainties to a level below 10\%.
Still, the heavy top limit, well justified
for energies below the top quark pair production threshold at about 345\,GeV, is problematic for the production of three Higgs bosons.

The phenomenology of triple Higgs production, in particular its sensitivity
to the quartic self-coupling and to beyond-the-Standard-Model~(BSM) effects,
has been explored in a series of
studies~\cite{Kilian:2017nio,Fuks:2017zkg,Liu:2018peg,Papaefstathiou:2020lyp,Stylianou:2023tgg,Biermann:2024oyy,Haisch:2025pql,Dong:2025lkm,Papaefstathiou:2025meh,Naskar:2026rhi,Braathen:2026jfg,Procter:2026rlk,Merianos:2971199}.
These works have investigated the interplay between the trilinear and quartic
couplings in the total cross section and differential distributions, and have
shown that deviations from SM coupling values can significantly enhance the
rate, making $gg\to HHH$ a potentially promising channel in BSM
scenarios.  A comprehensive overview of the status
and experimental prospects a few years ago, including benchmark scenarios and recommended
signal and background estimates, is provided by the triple Higgs production
White Paper~\cite{Abouabid:2024gms}.

%\GH{Say something about MC}

The description of $gg \to HHH$ within a systematic EFT framework has
attracted increasing attention recently, EFT studies of triple Higgs
production have been presented in
Refs.~\cite{Anisha:2024ljc,Fuks:2025gjv,Domenech:2025gmn,Panizzi:2025sya,Ding:2026qto,Campbell:2026oae}.
These studies have highlighted that the quartic coupling modifier $\kappa_4$
and the trilinear modifier $\kappa_3$ are strongly correlated in the total
rate, and that differential distributions, in particular the invariant masses
of the double and triple Higgs system, can be useful to disentangle the two.
However, these studies so far either adhered to the $\kappa$-framework or included only a subset of operators that would contribute to a strict EFT power counting scheme.

In this paper, we present a calculation of $gg\to HHH$, combining the one-loop predictions in the SM with 
Higgs Effective Field Theory (HEFT), including for the first time the full set of eight HEFT operators relevant at
leading chiral order.  This allows us to study the impact of effective gluon-Higgs couplings in combination with anomalous Higgs-top and Higgs self-couplings. We study the effects of individual operators as well as benchmark points that produce characteristic shapes of the triple Higgs invariant mass spectrum which would not occur upon variations of individual Wilson coefficients.
Furthermore, we study Higgs decay and parton shower effects within the HEFT framework, enabled by a new interface between \textsc{GoSam}-3~\cite{Braun:2025afl} and \textsc{Sherpa-3}~\cite{Sherpa:2024mfk}. 

The structure of this paper is as follows. In 
Section~\ref{sec:heft}, we briefly introduce the HEFT framework and derive the corresponding effective Lagrangian relevant to triple Higgs production.
Section~\ref{sec:setup} is dedicated to the description of the tools and the setup used for the calculation. We also describe in detail our validation against the results of Ref.~\cite{Campbell:2026oae}.
In Section~\ref{sec:results} we present our phenomenological results, before we conclude in Section~\ref{sec:conclusions}.
In an appendix we provide the parametrisation of the HEFT cross section as a polynomial in terms of 62 combinations of Wilson coefficients.

\section{Triple Higgs Production in the Higgs Effective Field Theory}
\label{sec:heft}
Two effective field theories (EFTs) are commonly used in the context of New Physics searches the LHC, the Standard Model Effective Field Theory (SMEFT)~\cite{Buchmuller:1985jz,Grzadkowski:2010es,Brivio:2017vri,Isidori:2023pyp,Aebischer:2025qhh} and the Higgs Effective Field Theory (HEFT)~\cite{Feruglio:1992wf,Bagger:1993zf,Koulovassilopoulos:1993pw,Burgess:1999ha,Wang:2006im,Grinstein:2007iv,Alonso:2012px,Buchalla:2012qq,Buchalla:2013rka}. Both allow the parameterisation of BSM effects at current collider energies assuming any New Physics (NP) scale lives well above the electroweak scale. We work in the HEFT framework, which is also commonly known as Electroweak Chiral Lagrangian (EWChL) or non-linear EFT. In this theory, the Higgs boson and the electroweak Goldstone bosons are treated as independent, the Higgs boson is included as an electroweak singlet.
%and not grouped in a doublet structure. 
This results in the Lagrangian being organised in terms of the \emph{chiral dimension} $d_\chi$, which can be related to the loop order $L$ as $d_\chi = 2 L + 2$. The chiral dimension of an operator can be calculated from the chiral dimensions of its fundamental building blocks \cite{Buchalla:2013eza}\footnote{For a definition at cross section level, that also takes into account the phase space, we refer to Ref.~\cite{Brivio:2025yrr}.}:
\begin{equation}
    d_\chi(\varphi, H) = 0, \qquad d_\chi(\partial, X_{\mu\nu}, \bar{\psi}\psi, \kappa) = 1.
\end{equation}
Here $\varphi$ is a Goldstone boson, $H$ the Higgs boson, $X_{\mu\nu}$ a gauge boson field strength tensor and $\kappa$ a gauge or Yukawa coupling.  
Within this description, the leading order (LO) HEFT Lagrangian $\mathcal{L}_2$ is a superset of the SM Lagrangian and can be written as~\cite{Buchalla:2013rka}
\begin{equation}
    \begin{aligned}
        \mathcal{L}_2 = & - \frac{1}{4} G_{\mu\nu}^a G^{a, \mu\nu} - \frac{1}{4} W^I_{\mu\nu}W^{I,\mu\nu} - \frac{1}{4} B_{\mu\nu}B^{\mu\nu} \\
        &+ \frac{1}{2} (\partial_\mu H)(\partial^\mu H) - \frac{v^2}{4} \tr\left((D_\mu U)^\dag(D^\mu U)\right) \mathcal{F}\left(\eta\right) - V(\eta) \\
        &+ \vphantom{\frac{1}{2}}\bar{\psi} \left(i \slashed{D} - U \mathcal{M}\left(\eta\right) P_R - \mathcal{M}^\dag\left(\eta\right)U^\dag P_L\right) \psi\;,
    \end{aligned}
\end{equation}
where $G^a_{\mu\nu}$, $W^I_{\mu\nu}$ and $B_{\mu\nu}$ are the SM $SU(3)_C$, $SU(2)_L$ and $U(1)_Y$ field strength tensors and $\eta = H / v$, with the electroweak symmetry breaking scale $v$. In this notation, the fermions are compactly grouped into a single object $\psi = (u, d, \nu, e)$ with the covariant derivative
\begin{equation}
    D_\mu \psi = \left(\partial_\mu + i g_s T^a G^a_\mu + i g \tau^I W^I_\mu P_L + i g' B_\mu(Y_L P_L + Y_R P_R)\right)\psi.
\end{equation}
The electroweak symmetry breaking is realised through the Goldstone boson matrix $U$, which is defined as
\begin{equation}
    U = \exp\left(\frac{2i}{v}\tau^I \varphi^I\right), \quad D_\mu U = \partial_\mu U + i g \tau^I W^I_\mu U - i g' B_\mu U \tau^3,
\end{equation}
where $\tau^I$ are the generators of $SU(2)_L$. Since the Higgs field contributes no chiral dimension, it can appear to arbitrary powers in the so-called Flare-function $\mathcal{F}$, the mass matrix $\mathcal{M} = (\mathcal{M}_u, \mathcal{M}_d, \mathcal{M}_\nu,\mathcal{M}_e)$ and the Higgs potential $V$:
\begin{equation}
    \mathcal{F}\left(\eta\right) = 1 + \sum_{n = 1}^\infty f_n \eta^n, \quad \mathcal{M}_f(\eta) = m_f + \sum_{n=1}^\infty m_{f,n} \eta^n \quad \text{and} \quad V(\eta) = v^4 \sum_{n=2}^\infty V_n \eta^n.
\end{equation}
In addition to the LO HEFT Lagrangian, the process $gg\to HHH$ also receives contributions from the NLO HEFT Lagrangian $\mathcal{L}_4$. Since an operator with $d_\chi = 4$ contributes at one loop-order in the chiral counting, tree-level diagrams built with one such operator insertion contribute at the same order as the one-loop diagrams with an arbitrary number of $\mathcal{L}_2$ insertions. Two such operators from $\mathcal{L}_4$ can induce Higgs boson production in gluon fusion at tree-level \cite{Buchalla:2013rka}:
\begin{equation}
    \mathcal{L}_4 \supset \frac{g_s^2}{2} G^a_{\mu\nu}G^{a,\mu\nu} \mathcal{F}_G(\eta) + \frac{g_s^2}{2}\varepsilon^{\mu\nu\rho\sigma} G^a_{\mu\nu}G^a_{\rho\sigma} \mathcal{F}_{\tilde{G}}(\eta) \quad \text{where} \quad \mathcal{F}_X(\eta) = \sum_{n = 1}^\infty f_{X,n} \eta^n\;,
\end{equation}
where the second operator is CP-odd.

Considering now specifically triple-Higgs production in gluon fusion, only operators with up to three Higgs bosons contribute. Therefore, the full effective Lagrangian for the considered process can be written as
\begin{equation} \label{eq:L_eff}
    \begin{aligned}
    L_\mathrm{eff} \supset &-m_t \left(c_t \frac{H}{v} + c_{2t} \frac{H^2}{v^2} + c_{3t} \frac{H^3}{v^3}\right)\bar{t}t - c_3 \frac{m_H^2}{2v}H^3 - c_4 \frac{m_H^2}{8v^2}H^4 \\
    & + \frac{\alpha_s}{8\pi}\left(c_g \frac{H}{v} + c_{2g} \frac{H^2}{v^2} + c_{3g} \frac{H^3}{v^3}\right) G^a_{\mu\nu}G^{a,\mu\nu} \\ 
    & + \frac{\alpha_s}{4\pi} \left(\tilde{c}_g \frac{H}{v} + \tilde{c}_{2g} \frac{H^2}{v^2} + \tilde{c}_{3g} \frac{H^3}{v^3}\right) \varepsilon^{\mu\nu\rho\sigma}G^a_{\mu\nu}G^a_{\rho\sigma}\;,
    \end{aligned}
\end{equation}
where we have defined the coupling parameters $c_i$ such that
%slightly changed the parameterisation of the couplings \GH{changed compared to what?}
they match previous double-Higgs studies \cite{deFlorian:2021azd,Heinrich:2020ckp,Heinrich:2019bkc,Buchalla:2018yce}. With this parametrisation, the SM Lagrangian can be restored by setting $c_3 = c_4 = c_t = 1$ and all other parameters to zero.

\begin{figure}
    \centering
    \begin{subfigure}{0.19\linewidth}
        \includegraphics[width = \linewidth]{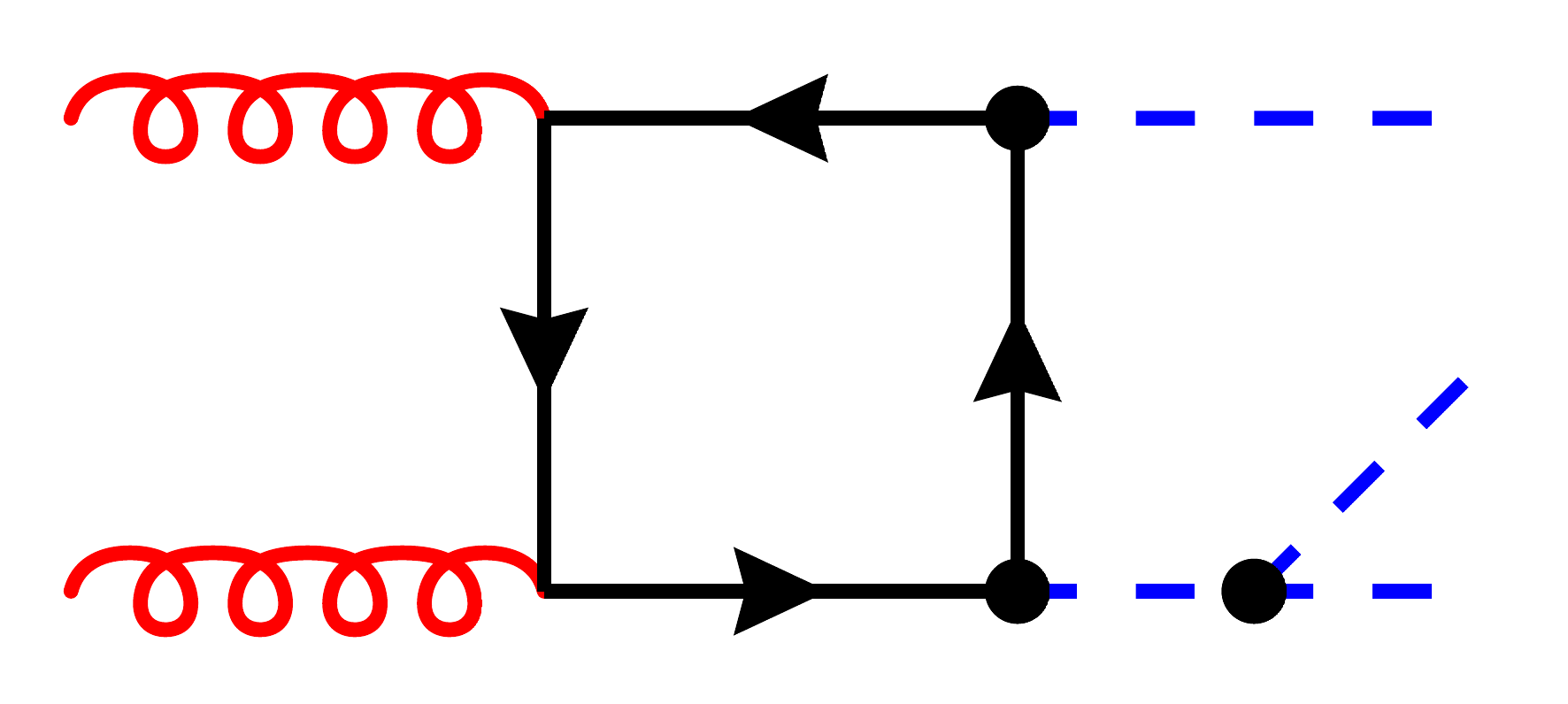}
    \end{subfigure}
    \begin{subfigure}{0.19\linewidth}
        \includegraphics[width = \linewidth]{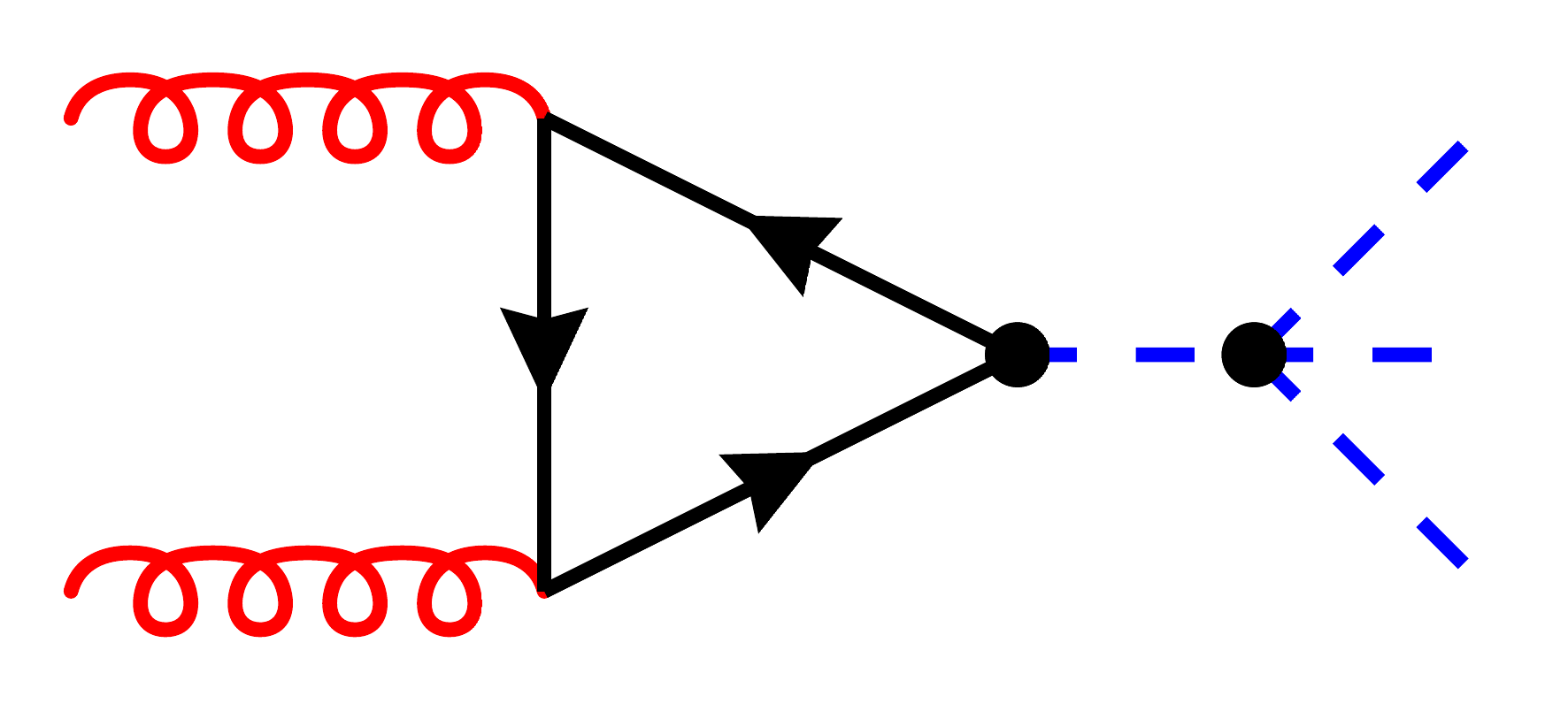}
    \end{subfigure}
    \begin{subfigure}{0.19\linewidth}
        \includegraphics[width = \linewidth]{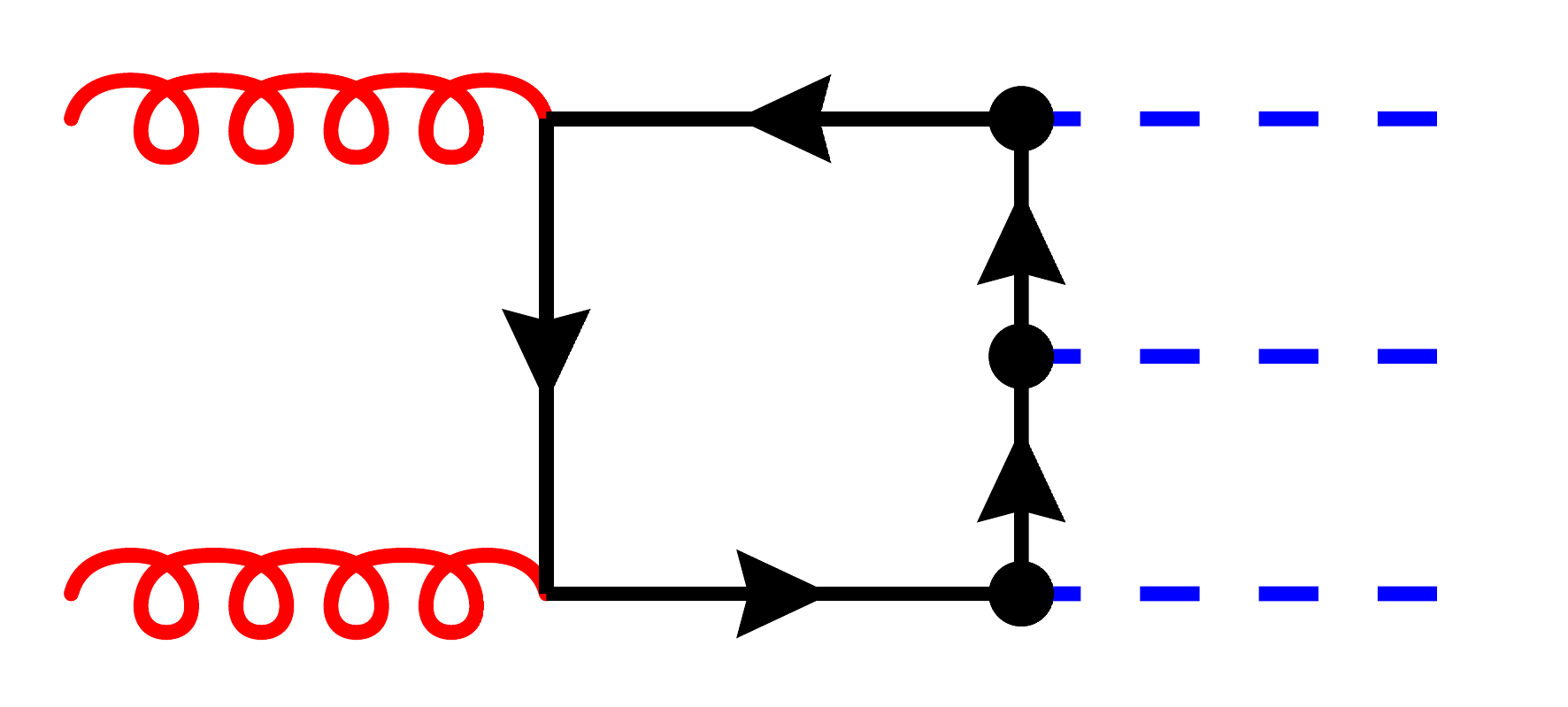}
    \end{subfigure}
    \begin{subfigure}{0.19\linewidth}
        \includegraphics[width = \linewidth]{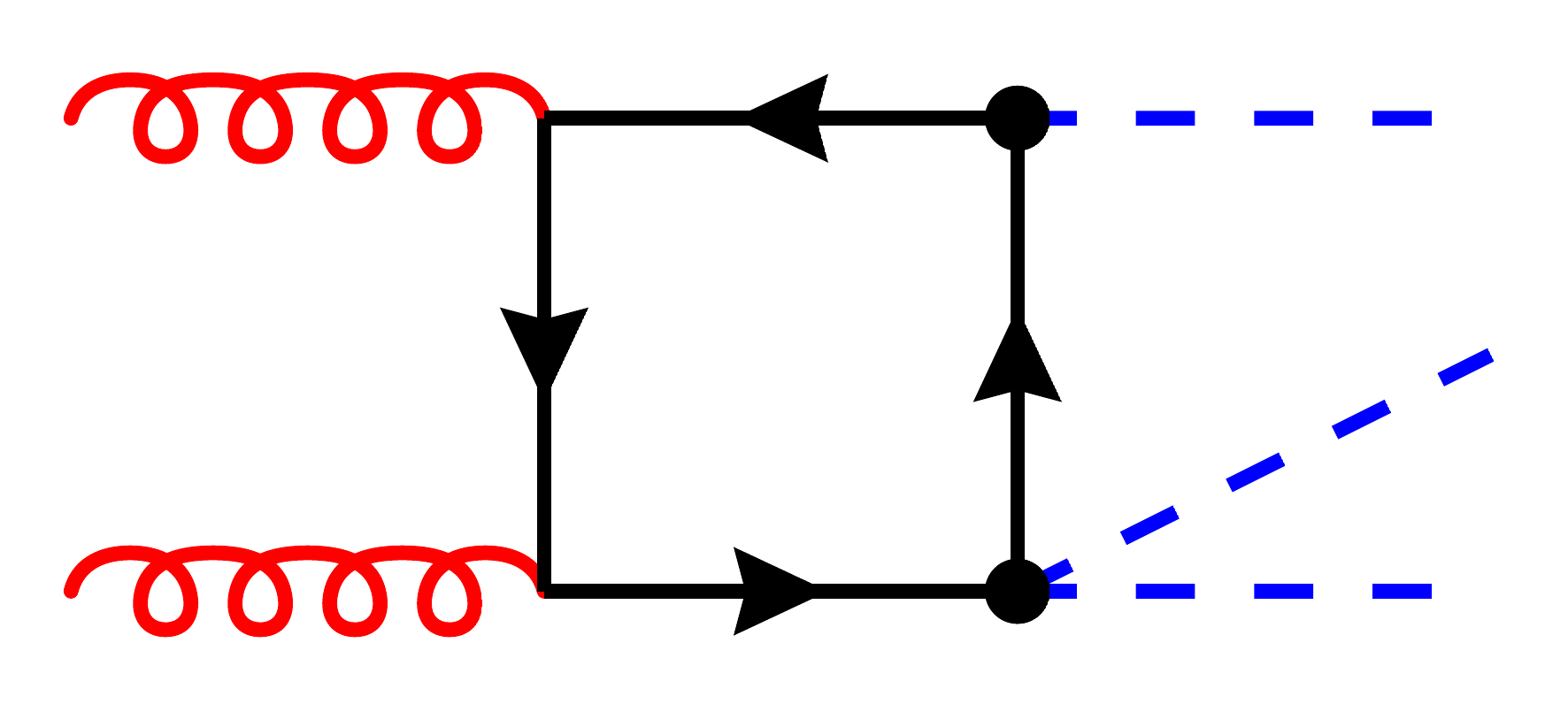}
    \end{subfigure}
    \begin{subfigure}{0.19\linewidth}
        \includegraphics[width = \linewidth]{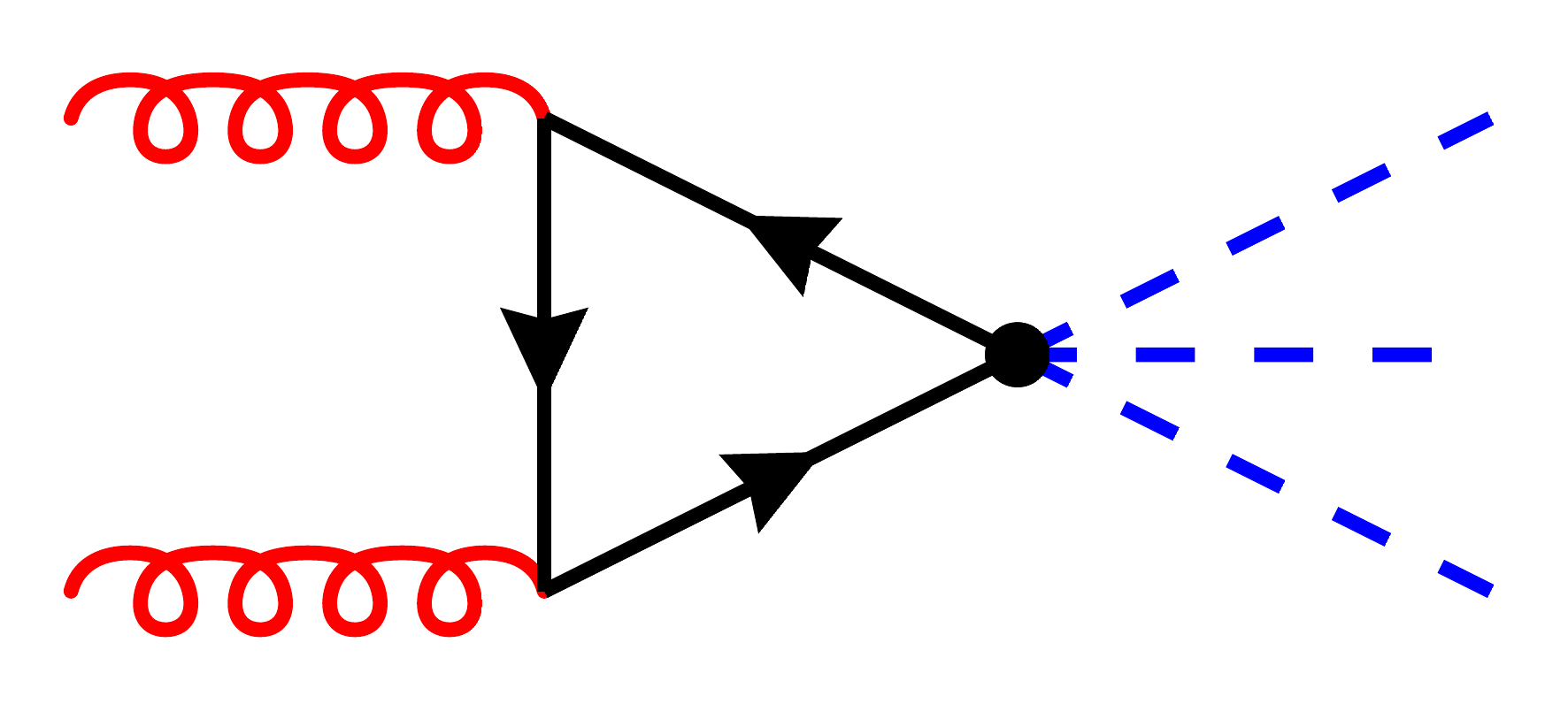}
    \end{subfigure}
    \begin{subfigure}{0.19\linewidth}
        \includegraphics[width = \linewidth]{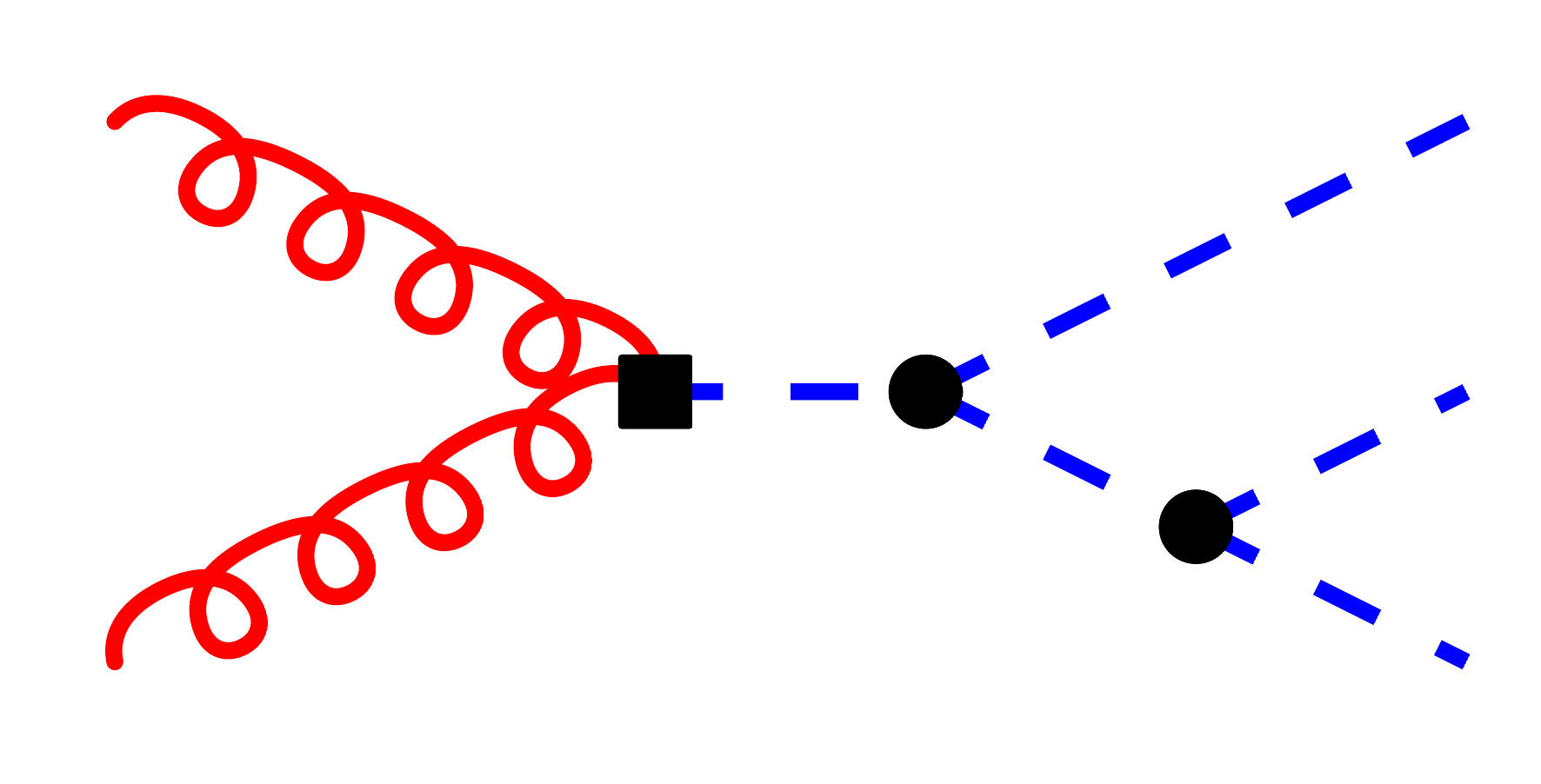}
    \end{subfigure}
    \begin{subfigure}{0.19\linewidth}
        \includegraphics[width = \linewidth]{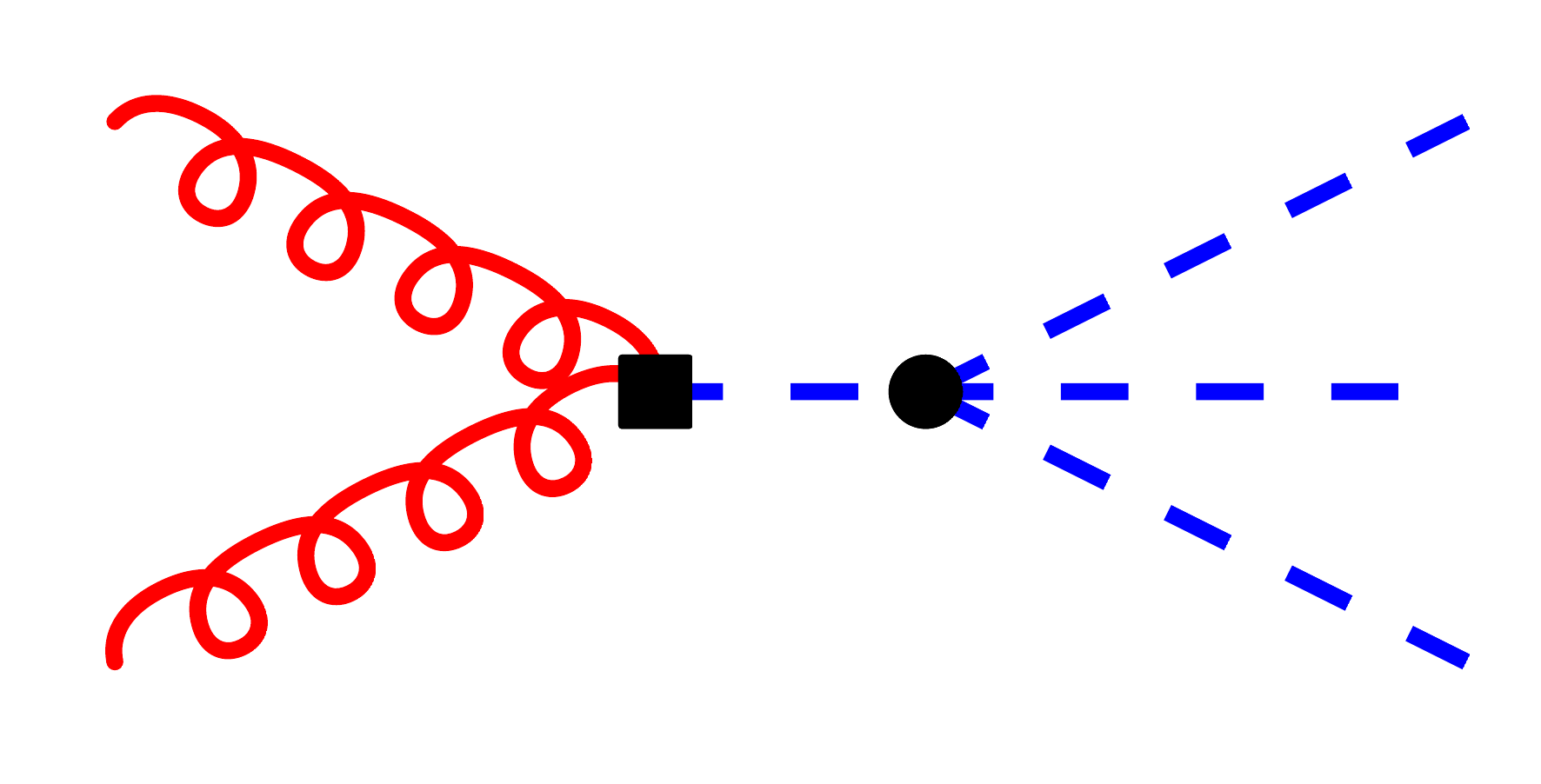}
    \end{subfigure}
    \begin{subfigure}{0.19\linewidth}
        \includegraphics[width = \linewidth]{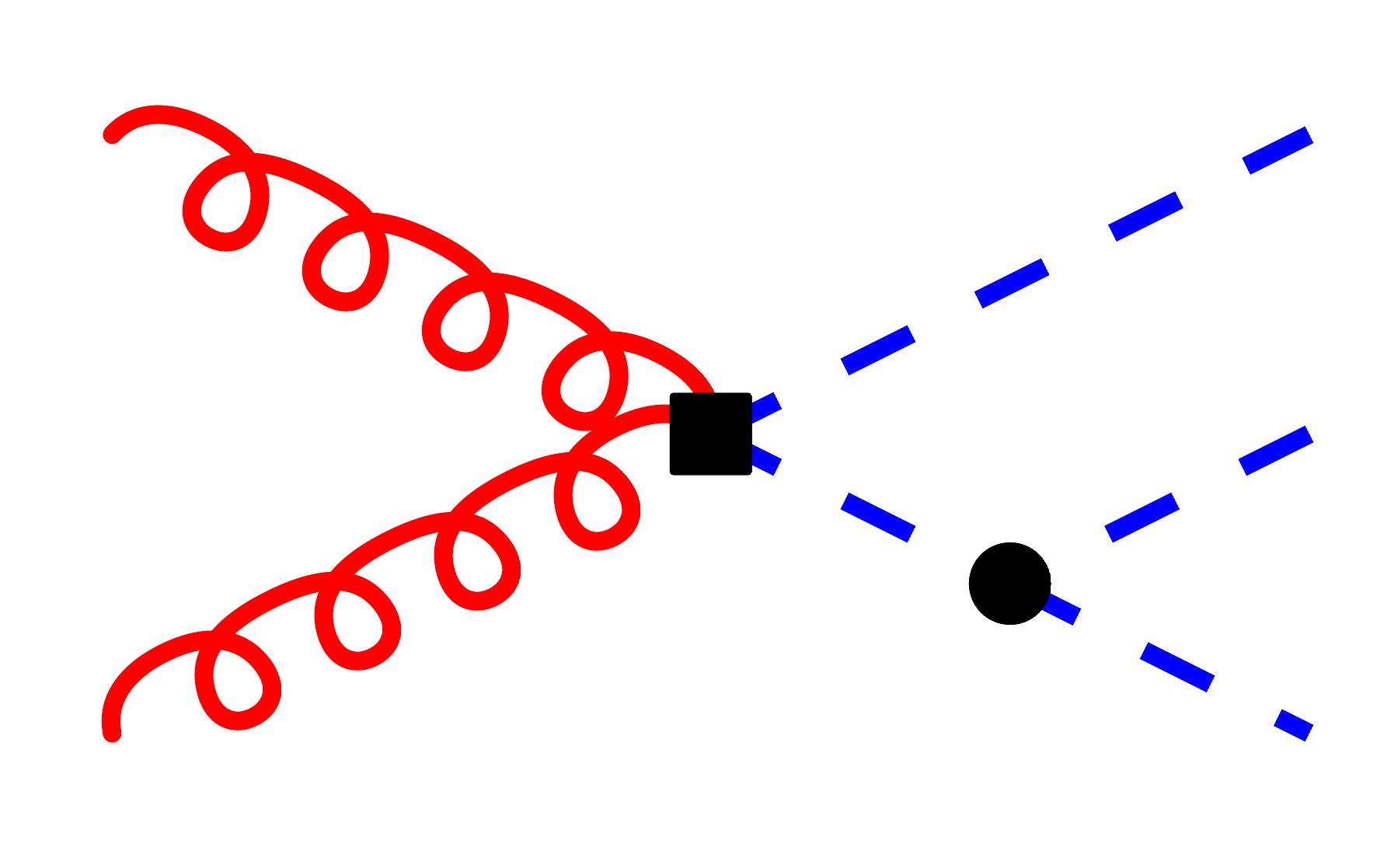}
    \end{subfigure}
    \begin{subfigure}{0.19\linewidth}
        \includegraphics[width = \linewidth]{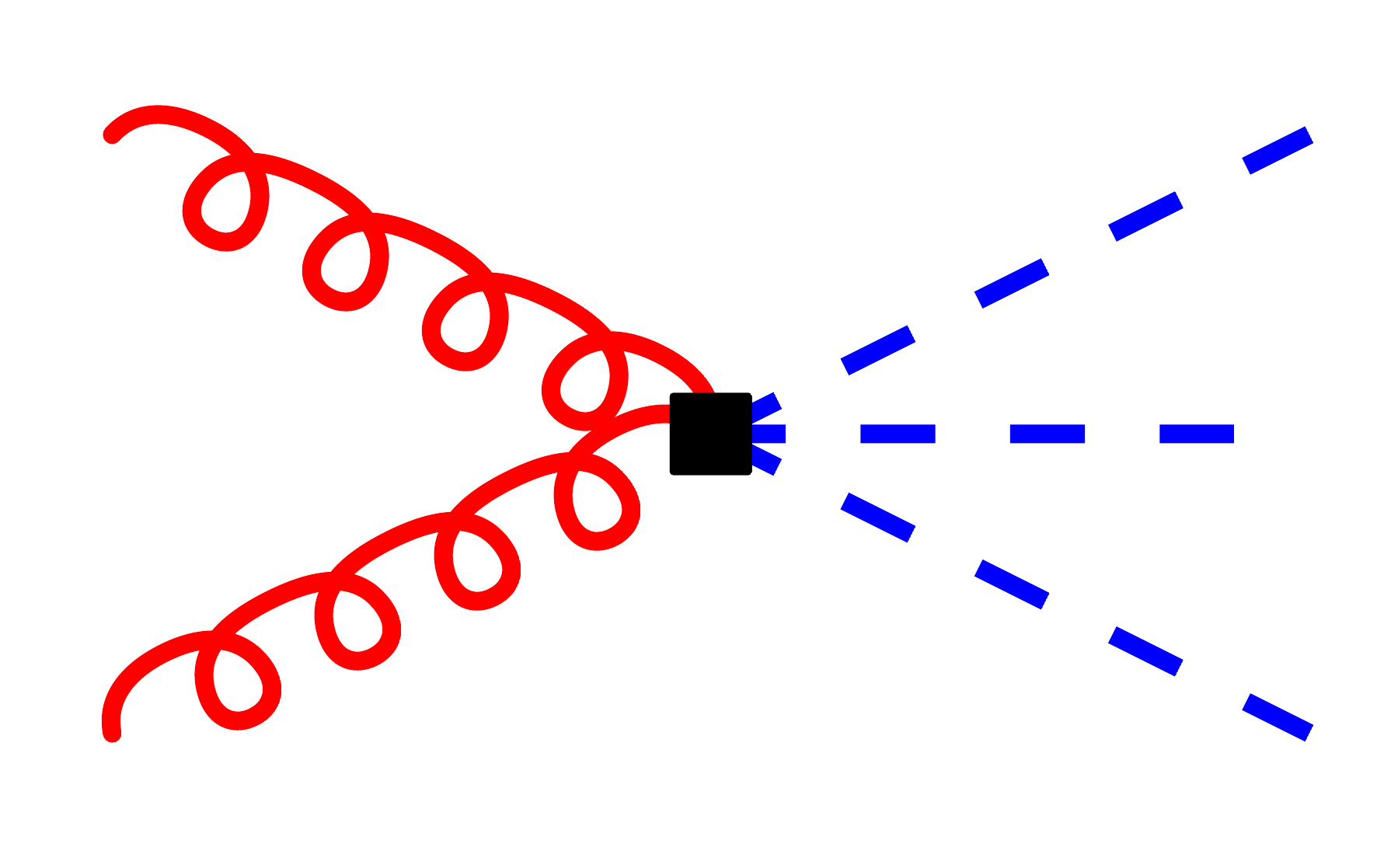}
    \end{subfigure}
    \caption{Selection of Feynman diagrams contributing to triple Higgs production in gluon fusion within the HEFT framework. Black squares on vertices indicate a coupling from $\mathcal{L}_4$, circles a coupling from $\mathcal{L}_2$.}
    \label{fig:diagrams}
\end{figure}

With the Lagrangian in equation \eqref{eq:L_eff}, a combination of tree-level diagrams with a single $\mathcal{L}_4$ insertion and arbitrary numbers of $\mathcal{L}_2$ insertions, as well as one-loop diagrams with only $\mathcal{L}_2$ insertions, will contribute to our calculation. This results in eight tree-level diagrams and 76 one-loop diagrams, a subset of which is shown in Fig.~\ref{fig:diagrams}. Due to both classes being effectively at the one-loop order, the amplitude is then calculated as 
\begin{equation}
    \left|\mathcal{M}\right|^2 = \left|\mathcal{M}_\mathrm{tree} + \mathcal{M}_\mathrm{one-loop}\right|^2.
\end{equation}

\section{Framework of the Calculation and Setup}
\label{sec:setup}
We compute our results with a general-purpose setup linking the one-loop provider \textsc{GoSam} \cite{Cullen:2011ac, GoSam:2014iqq, Braun:2025afl} and the Monte Carlo event generator \textsc{Sherpa} \cite{Gleisberg:2008ta,Sherpa:2019gpd,Sherpa:2024mfk}. This interface is based on the Binoth Les Houches Accord (BLHA) \cite{Binoth:2010xt,Alioli:2013nda} interface and was originally created between \textsc{GoSam}-1 and \textsc{Sherpa}-1 and subsequently used in various SM NLO computations \cite{vanDeurzen:2013rv,Hoeche:2013mua,vanDeurzen:2013xla,Cullen:2013saa,Heinrich:2013qaa,Greiner:2015jha}. For this study, we have updated the interface to the most recent versions of both codes, \textsc{GoSam}-3~\cite{Braun:2025afl} and \textsc{Sherpa}-3~\cite{Sherpa:2024mfk}. As \textsc{GoSam}-3 can generate amplitudes including EFT operators in an automated way, this opens the door to precision EFT studies within the \textsc{Sherpa} framework, including loop induced processes. It complements \sherpa's internal interface to UFO models \cite{Hoche:2014kca}, that has been used to study processes in SMEFT at LO \cite{Biekotter:2020flu,Biekotter:2021int,Banerjee:2024eyo}, and NLO implementations of SMEFT corrections dedicated to specific processes \cite{Maskos:2026wcy}. The interface can make use of \sherpa's ability to reweigh matrix elements on-the-fly \cite{Bothmann:2016nao} to store multiple weights per event in the associated \hepmc record \cite{Buckley:2019xhk,Bothmann:2022pwf}, corresponding to different EFT parameter points. This means one sample, needing only one run of the parton shower, non-perturbative modelling and potentially further processing outside \sherpa, can be generated for multiple parameter points of interest at once. \sherpa's interface to \rivet \cite{Bierlich:2019rhm,Bierlich:2024vqo} automatically handles these weights in analyses such that histograms for all chosen points are automatically available.

In addition to the use as matrix element generator for loop induced processes,  the interface allows \textsc{GoSam} to be used as a generator for tree-level matrix elements, for example as input to \textsc{Sherpa}'s multijet merging using the CKKW merging method \cite{Catani:2001cc} and to provide virtual amplitudes to the internal matrix element generators \amegic \cite{Krauss:2001iv,Gleisberg:2007md} and \comix \cite{Gleisberg:2008fv} for matching using the \textsc{MCatNLO}~\cite{Frixione:2002ik} method as implemented in \sherpa \cite{Hoeche:2011fd}, and hence for \sherpa's \MEPSatLO \cite{Hoeche:2009rj} and \MEPSatNLO \cite{Hoeche:2012yf} methods.
Events are showered with \textsc{Sherpa}'s default \CSS \cite{Schumann:2007mg} based on Catani-Seymour dipoles \cite{Catani:1996vz}, or alternatively with the \alaric \cite{Herren:2022jej,Assi:2023rbu,Hoche:2024dee} parton shower. 
Subsequently, events can be hadronised using the cluster hadronisation model \cite{Webber:1983if} as implemented in \textsc{Sherpa}~\cite{Winter:2003tt,Chahal:2022rid}, or alternatively the Lund model using an interface to \textsc{Pythia}-8 \cite{Bierlich:2022pfr}. 

The squared matrix element for the process $gg\to hhh$, consisting of tree-level and one-loop diagrams contributing up to chiral dimension four to this process, is provided by \textsc{GoSam}. The model is supplied to \textsc{GoSam} in the UFO-2 format \cite{Degrande:2011ua,Darme:2023jdn}, based on which \textsc{FeynGraph} \cite{feyngraph} produces all Feynman diagrams. These are then translated to analytical expressions and simplified via \textsc{Form} \cite{Vermaseren:2000nd,Kuipers:2012rf,Davies:2026cci}. Finally, the integral reduction is performed numerically by \textsc{Ninja} \cite{vanDeurzen:2013saa,Peraro:2014cba} and the master integrals are provided by \textsc{OneLOop} \cite{vanHameren:2010cp}. Notably, the correct combination of tree-level and one-loop contributions is performed automatically by \textsc{GoSam} through the \texttt{EFTcount} configuration option introduced in \textsc{GoSam}-3. 

All configuration files necessary to run this setup, including the UFO model, the \textsc{Rivet} analysis and the \textsc{Sherpa} input files are provided in a public repository\footnote{\url{https://github.com/Jens-Braun/ggHHH-HEFT}}.

\subsection{Validation}
\label{ssec:valid}
We validate our setup in multiple ways, beginning with a validation of our general setup and our UFO model. For this purpose, we first reduce the number of produced Higgs bosons and consider double Higgs production in gluon fusion. The number of contributing HEFT couplings then reduces to five: $c_3$, $c_t$, $c_{2t}$, $c_g$ and $c_{2g}$. We produce cross sections and differential distributions for seven benchmark points defined in \cite{Heinrich:2020ckp} and compare our results to the ones produced by the \textsc{Powheg} implementation described therein.

\begin{figure}
    \centering
    \begin{subfigure}{0.49\linewidth}
        \includegraphics[width=\linewidth, page = 1]{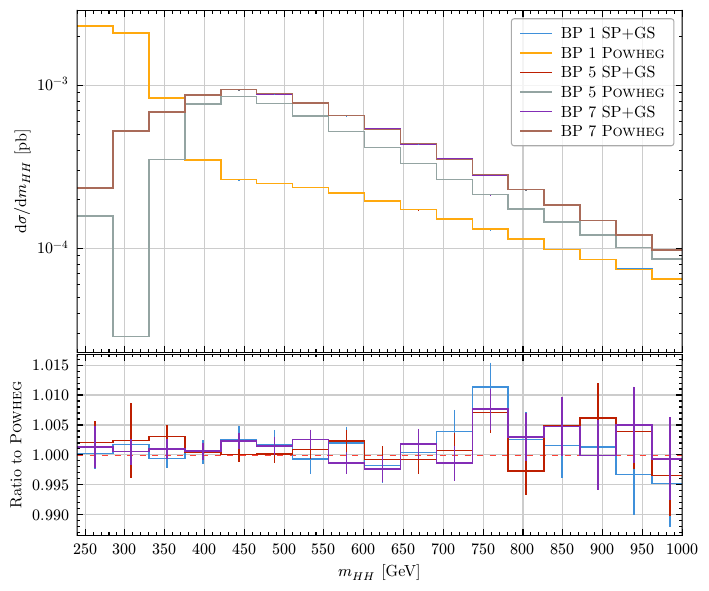}
    \end{subfigure}
    \begin{subfigure}{0.49\linewidth}
        \includegraphics[width=\linewidth, page = 2]{figures/ggHH_Comparison.pdf}
    \end{subfigure}
    \caption{Higgs boson pair invariant mass (left) and transverse momentum of any Higgs boson (right) for benchmark points 1, 5 and 7 from table \ref{tab:powheg_comparison}.}
    \label{fig:powheg_comparison}
\end{figure}

The total (LO) cross sections for these points are summarised in Tab.~\ref{tab:powheg_comparison}, which shows perfect agreement between the tools within the Monte Carlo uncertainty. Fig.~\ref{fig:powheg_comparison} shows invariant mass and transverse momentum distributions for three of these benchmark points, which again agree within the uncertainty. 

\begin{table}
    \centering
    \begin{tabular}{c | c c c c c | S[table-format=1.4] S[table-format=1.6] c}
        \toprule
         BP & $c_3$ & $c_t$ & $c_{2t}$ & $c_g$ & $c_{2g}$ & {\textsc{Powheg} [fb]} & {SP+GS [fb]} & Ratio \\
         \midrule
         1 & 3.94 & 0.94 & -1/3 & 0.5  & 1/3  & 0.1227(2)  & 0.12278(9) & 1.000(2) \\
         \rowcolor[gray]{0.9} 2 & 6.92 & 0.61 & 1/3  & 0.0  & -1/3 & 0.0850(1)  & 0.08506(6) & 0.999(2) \\
         3 & 2.21 & 1.05 & -1/3 & 0.5  & 0.5  & 0.0878(1)  & 0.87710(6) & 1.000(2) \\
         \rowcolor[gray]{0.9} 4 & 2.79 & 0.61 & 1/3  & -0.5 & 1/6  & 0.03083(4) & 0.03083(2) & 1.000(2) \\
         5 & 3.95 & 1.17 & -1/3 & 1/6  & -0.5 & 0.1003(2)  & 0.10006(8) & 1.002(2) \\
         \rowcolor[gray]{0.9} 6 & 5.68 & 0.83 & 1/3  & -0.5 & 1/3  & 0.08717(1) & 0.08707(6) & 1.001(2) \\ 
         7 & -0.1 & 0.94 & 1    & 1/6  & 1/6  & 0.1268(2)  & 0.12696(9) & 0.999(2) \\
         \bottomrule
    \end{tabular}
    \caption{Total cross section for $gg \rightarrow HH$ at leading order calculated with our Sherpa+GoSam setup and the \textsc{Powheg} setup of \cite{Heinrich:2020ckp} for seven benchmark points. The uncertainty denotes the Monte Carlo integration error.}
    \label{tab:powheg_comparison}
\end{table}

Secondly, we test our results in double-Higgs production when including the CP-odd couplings $\tilde{c}_g$ and $\tilde{c}_{2g}$ against the \textsc{Hpair} implementation presented in \cite{Grober:2017gut}. For this purpose, we compare the total cross section for several combinations of the couplings and find excellent agreement within the statistical uncertainty between the two setups.

Finally, we return to triple Higgs production and compare our results against another recent calculation \cite{Campbell:2026oae}, which contains a subset of the operators we include. First, we compare the total cross section by randomly selecting coupling values to evaluate the parameterisation of \cite{Campbell:2026oae} on and compared this total cross section prediction with our parameterisation given in section \ref{ssec:xsec}. We see good agreement within the statistical expectation over the entire considered coupling parameter space. Secondly, we reproduce the invariant mass distribution of the triple-Higgs system for parameter points given in Fig. 7 of \cite{Campbell:2026oae} and again find excellent agreement between our setups.

\section{Phenomenological Results}
\label{sec:results}
In this chapter, we consider the LHC at a centre-of-mass energy of $\sqrt{s} = 14~\mathrm{TeV}$. We choose the \texttt{alphaMZ} scheme as our electroweak input scheme, in which the input parameters are defined as
\begin{equation}
    \alpha^{-1}(M_Z) = 132.048, \qquad M_Z = 91.1876~\mathrm{GeV} \quad \text{and} \quad M_W = 80.3625~\mathrm{GeV},
\end{equation}
resulting in $v = 246.218~\mathrm{GeV}$. The masses of the remaining appearing particles are set to 
\begin{equation}
    m_t = 173~\mathrm{GeV} \quad \text{and} \quad m_H = 125~\mathrm{GeV}
\end{equation}
and we do not include effects of a finite top width or bottom quarks. We use a running $\alpha_s$ from the \texttt{PDF4LHC21\_40} parton distribution functions (PDFs)~\cite{PDF4LHCWorkingGroup:2022cjn}, interfaced through \textsc{Lhapdf}~\cite{Buckley:2014ana}. Finally, we choose the central renormalisation and factorisation scale \begin{equation}
    \mu_0 = \frac{1}{2} m_{HHH}.
\end{equation}

We do not include the couplings $\tilde{c}_g$, $\tilde{c}_{2g}$ and $\tilde{c}_{3g}$ in this chapter, since we found the impact of the linear CP-odd term to be a less-than-per-mille contribution compared to the quadratic term and therefore focus solely on the CP-even operators.

\subsection{Total Cross Section}
\label{ssec:xsec}
We begin our study by first considering the total cross section. We orient the allowed coupling ranges at recent experimental bounds where available \cite{CMS:2026zxu,CMS:2026xkc,CMS:2026nuu,ATLAS:2026pdi,CMS:2025ngq,ATLAS:2024xcs,ATLAS:2024ish}, using conservative intervals:
\begin{equation} \label{eq:bounds_1}
    c_3 \in [-2, 7], \quad c_4 \in [-200, 200], \quad c_t \in [0.8, 1.2], \quad c_g \in [-1, 1].
\end{equation}
The remaining ranges are chosen such that the total cross section relative to the SM does not significantly exceed the recent bounds of 588 and 760 of CMS \cite{CMS:2026xkc} and ATLAS \cite{ATLAS:2024xcs}, respectively:
\begin{equation} \label{eq:bounds_2}
    c_{2g} \in [-8, 8], \quad c_{3g} \in [-1.5, 1.5], \quad c_{2t} \in [-6, 6], \quad c_{3t} \in [-3, 3].
\end{equation}

We parameterise the total cross section as a polynomial in the anomalous couplings, which we give in Appendix \ref{sec:app:parameterisation}. This polynomial has 62 coefficients, which we determine by evaluating the total cross section for 1200 randomly sampled coupling values and fitting equation \eqref{eq:xsec} to the results using \texttt{iminuit} \cite{James:1975dr, iminuit}\footnote{The numerical values and errors of the polynomial coefficients are included in machine readable form in the accompanying repository at \url{https://github.com/Jens-Braun/ggHHH-HEFT}.}.
All following cross sections are normalised to the SM cross section, which we find to be 
\begin{equation}
    \sigma_\mathrm{SM} = 50.937(6)~\mathrm{ab}
\end{equation}
in our setup.

\begin{figure}
    \centering
    \includegraphics[width=0.8\linewidth,page = 1]{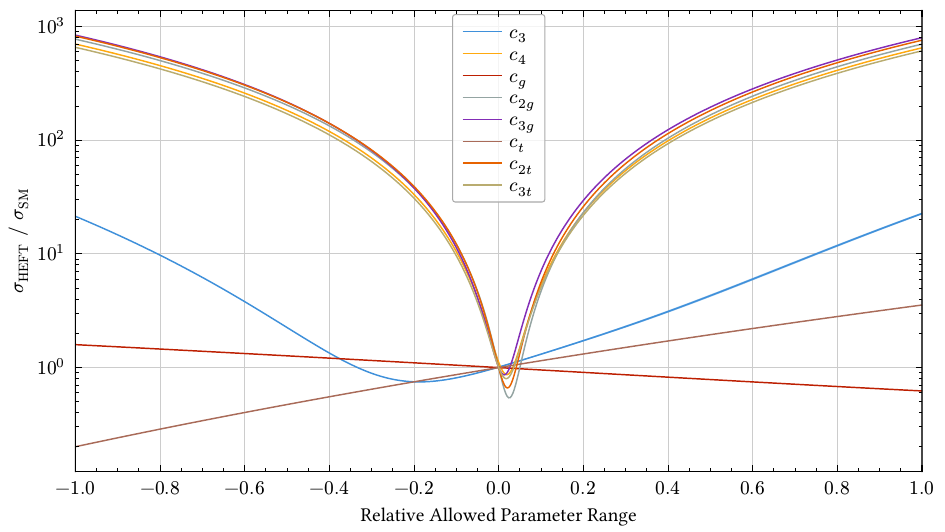}
    \caption{Total cross section relative to the SM value when exactly one coupling is varied and all others are fixed to the SM value. The parameters are varied within the ranges given in equations \eqref{eq:bounds_1} and \eqref{eq:bounds_2}.}
    \label{fig:xsec_single}
\end{figure}

Equipped with this parameterisation, we first consider the overall range of the total cross section variation within our coupling ranges. The total cross section rises rapidly with all experimentally unconstrained couplings, quickly reaching the current signal strength bounds. In the opposite direction, all couplings also allow for a reduction of the cross section relative to the total cross section, which can reach an overall reduction of up to $95\%$. This behaviour is also visible in Fig.~\ref{fig:xsec_single}, which shows the variation of exactly one coupling at a time. The remaining couplings are set to their SM value. The partitioning of the couplings into two classes is immediately visible here. The influence of $c_3$, $c_g$ and $c_t$ on the total cross section is more moderate compared to the other couplings due to the stronger bounds on these couplings from single and double Higgs measurements. In contrast to this, the other couplings modify the cross section by several orders of magnitude. 

\begin{figure}
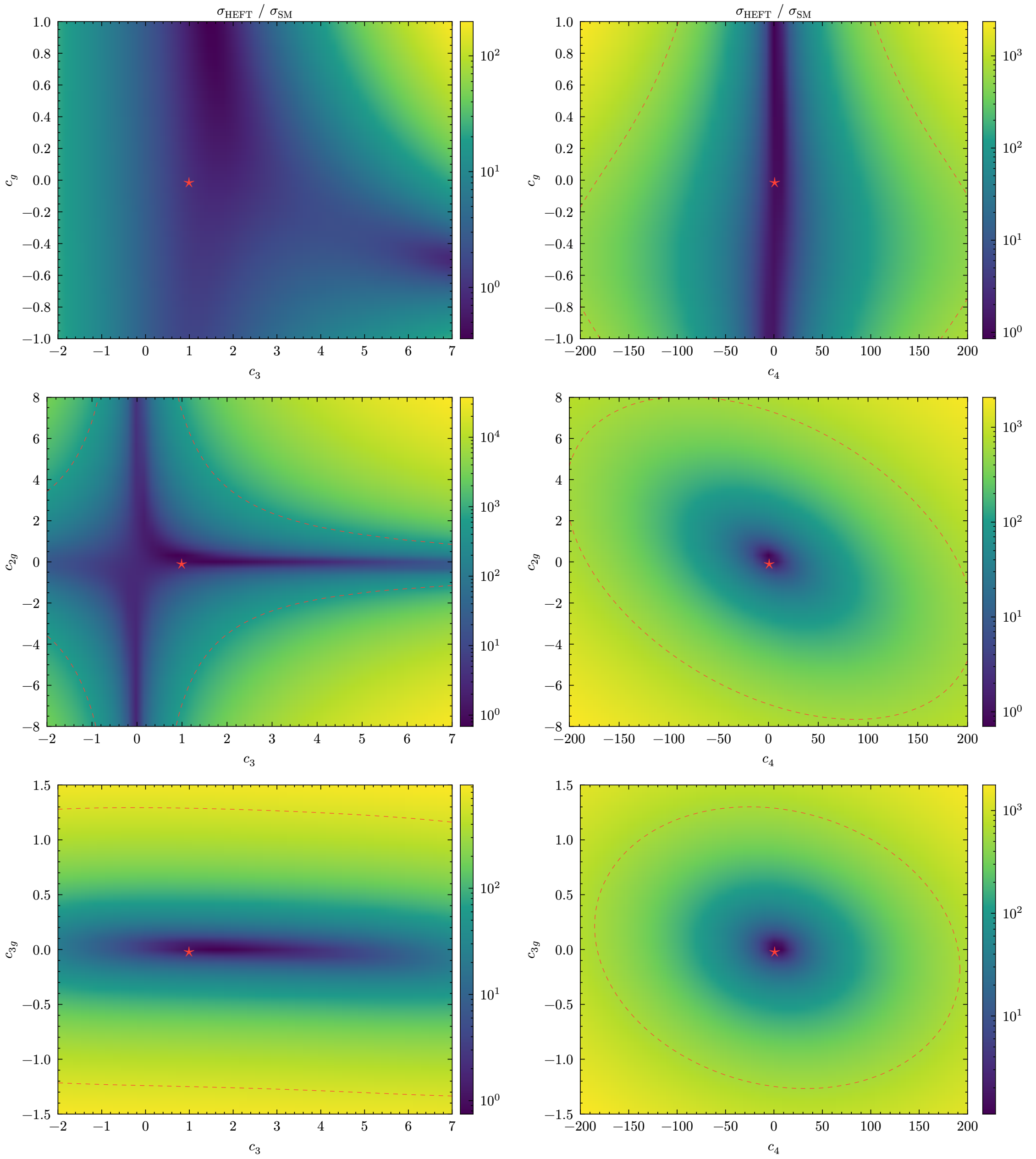

    \centering
    \begin{subfigure}[l]{0.49\linewidth}
       \includegraphics[width=\linewidth, page=3]{figures/plot.pdf} 
    \end{subfigure}
    \begin{subfigure}[r]{0.49\linewidth}
       \includegraphics[width=\linewidth, page=6]{figures/plot.pdf} 
    \end{subfigure}
    \begin{subfigure}[l]{0.49\linewidth}
       \includegraphics[width=\linewidth, page=4]{figures/plot.pdf} 
    \end{subfigure}
    \begin{subfigure}[r]{0.49\linewidth}
       \includegraphics[width=\linewidth, page=7]{figures/plot.pdf} 
    \end{subfigure}
    \begin{subfigure}[l]{0.49\linewidth}
       \includegraphics[width=\linewidth, page=5]{figures/plot.pdf} 
    \end{subfigure}
    \begin{subfigure}[r]{0.49\linewidth}
       \includegraphics[width=\linewidth, page=8]{figures/plot.pdf} 
    \end{subfigure}
    \caption{Total cross section relative to the SM value when exactly two couplings are varied and all others are fixed to the SM value. The red star indicates the SM value and the red contour marks the $\sigma_\mathrm{HEFT} / \sigma_\mathrm{SM} = 588$ boundary.}
    \label{fig:xsec_double}
\end{figure}

In this visualisation, the couplings $c_4$, $c_{2g}$, $c_{3g}$, $c_{2t}$ and $c_{3t}$ seem to behave very similarly. However, this merely is a consequence of only varying a single coupling at a time. This is visible in Fig.~\ref{fig:xsec_double}, which shows the cross section when $c_3$ or $c_4$ and $c_g$, $c_{2g}$ or $c_{3g}$ are varied at the same time. Here, a clear difference is observable in the contour shapes between the gluon couplings. Additionally, all of these plots show a strong dependence on the contact interactions, potentially significantly affecting the determination of coupling bounds when not considered.

\begin{figure}
    \centering
   \includegraphics[width=0.5\linewidth, page=2]{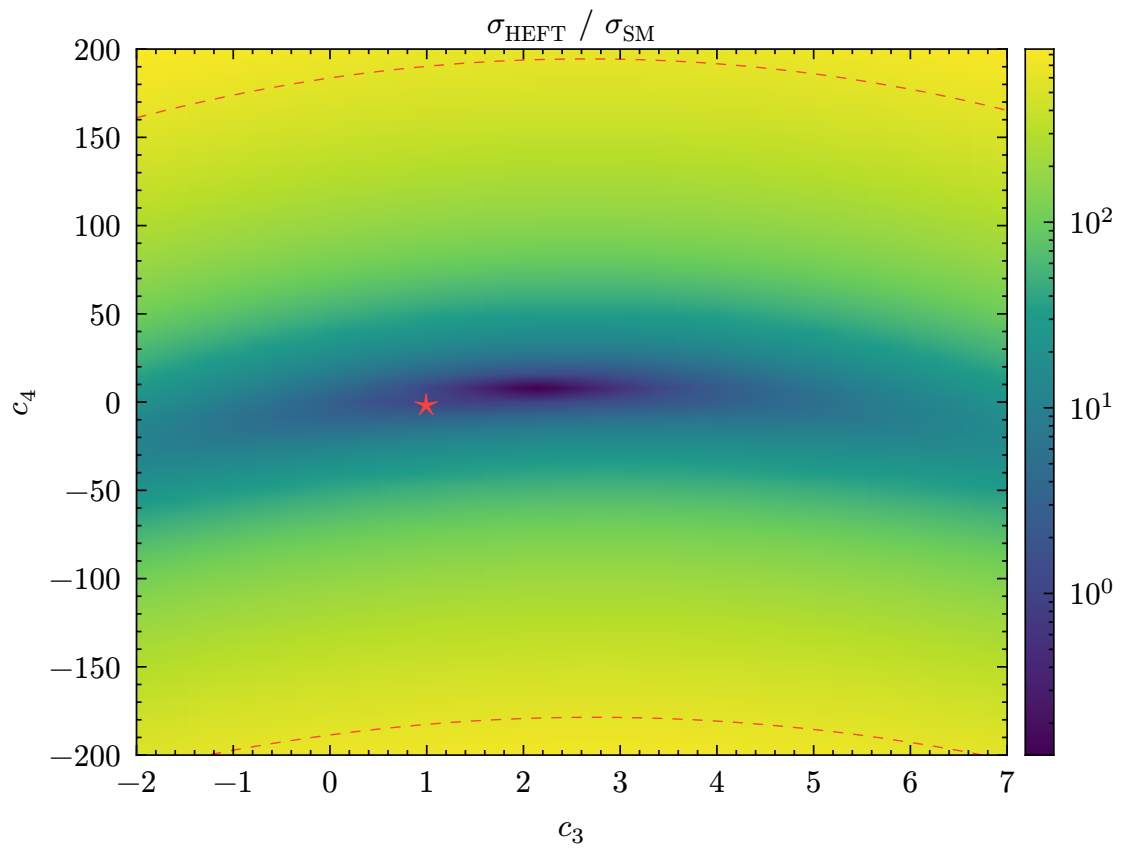}
    \caption{Total cross section relative to the SM value when varying the Higgs self couplings. The red star indicates the SM value and the red contour marks the $\sigma_\mathrm{HEFT} / \sigma_\mathrm{SM} = 588$ boundary.}
    \label{fig:xsec_c3c4}
\end{figure}

\subsection{On-Shell Higgs Distributions}
\label{ssec:stable}
\begin{figure}
    \centering
    \begin{subfigure}[l]{0.49\linewidth}
        \includegraphics[width=\linewidth]{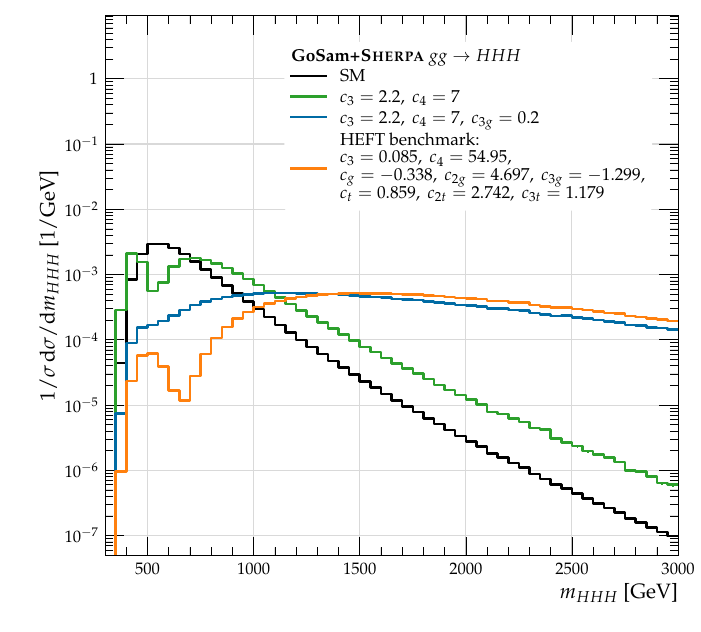}
    \end{subfigure}
    \begin{subfigure}[r]{0.49\linewidth}
        \includegraphics[width=\linewidth]{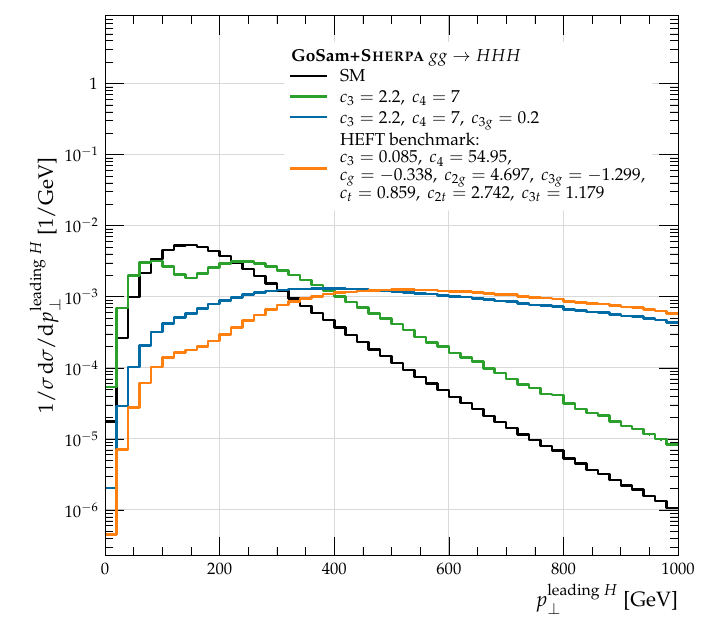}
    \end{subfigure}
    \caption{Invariant-mass distribution of the triple-Higgs system (left) and transverse momentum distribution of the Higgs boson with the largest transverse momentum (right) for the SM and three benchmark points.}
    \label{fig:onshell_dist}
\end{figure}

We turn our attention to differential distributions at matrix-element level, i.e.\ without taking into account Higgs-boson decays or initial-state radiation.
Fig.~\ref{fig:onshell_dist} shows the invariant mass of the triple-Higgs system as well as the transverse momentum of the leading Higgs boson.
To compare the shape change introduced by non-zero HEFT parameters compared to the SM point, we show normalised distributions of three different setups.

We first show the point $c_3=2.2$ and $c_4=7$ with all other couplings fixed to their SM values, located near the minimum of the total cross section in the $c_3$-$c_4$-plane visible in Fig.~\ref{fig:xsec_c3c4}. These values are chosen to reproduce the negative interference \cite{deFlorian:2019app} that produces a sharp dip around $m_{HHH} \approx 550$ GeV, where the peak of the SM distribution is located. We observe precisely this peak-dip-structure in the first point in Fig.~\ref{fig:onshell_dist}.
However, this specific feature in this configuration is not particularly stable with respect to other HEFT couplings, which we illustrate by also setting $c_{3g} = 0.2$. With this addition, the dip is fully compensated for by the increase in high-$m_{HHH}$ events caused by the contact interaction. Nonetheless, similar distribution shapes can be found with contributions from all HEFT couplings, for which we show the benchmark point
\begin{equation*}
  \begin{aligned}
    c_3 &= 0.085, & c_{4} &= 54.95, & c_g &= -0.338, & c_{2g} &= 4.697, \\
    c_{3g} &= -1.299, & c_t &= 0.859, & c_{2t} &= 2.742, & c_{3t} &= 1.179\,. \\
  \end{aligned}
\end{equation*}
found in a random search. All parameter points also lead to a significant enhancement of the large $m_{HHH}$ tail, by several orders of magnitude in the case of the points including the contact interactions.

The transverse momentum of the leading Higgs boson mirrors this behaviour.
For the first two points, a dip around the SM peak is produced by the same negative interference and removed again when including the contact interaction.
For the third point, it also distorts the shape, although it does not lead to a minimum in this specific setup.
All benchmark points enhance the tail significantly, the points with contact interactions again more than the point where only $c_3$ and $c_4$ are different from the SM.

\subsection{Decays and Parton Shower}
\label{ssec:decay_ps}
Next, we analyse events with \sherpa's parton showers included, and with all Higgs bosons decaying into bottom-anti-bottom quark pairs.
To assess the dependence on the shower algorithm, we produce samples with the \CSS and \alaric parton showers.
\alaric uses a global recoil and we use a dedicated scheme \cite{Hoche:2025gsb,Hoche:2026dup} that keeps the recoil within the decay system of a Higgs boson.
We cluster the resulting partonic final state into anti-$k_t$ \cite{Cacciari:2008gp} jets with radius $R=0.4$.
Jets are required to have $p_\perp > 25$~GeV and to lie at central rapidity, $|\eta| < 2.5$.
We use truth information to tag b-quarks as originating from the Higgs decays, and require that all 6 quarks are contained in accepted jets and that each jet contains at most one tagged quark.
Any event not satisfying these requirements is discarded.
This means there is no ambiguity in identifying exactly two jets per Higgs boson as originating from its decay products.
In an experimental setup, one would have to make an algorithmic choice about how to match jets likely originating from the same Higgs decay.
We also shower events with stable Higgs bosons, where all final-state partons are produced from the shower.
Histograms from this sample are scaled by the flat acceptance of the decayed sample,
$0.245$ in the case of \CSS and $0.240$ for \alaric.

\begin{figure}[!htb]
    \centering
    \includegraphics[width=0.49\linewidth]{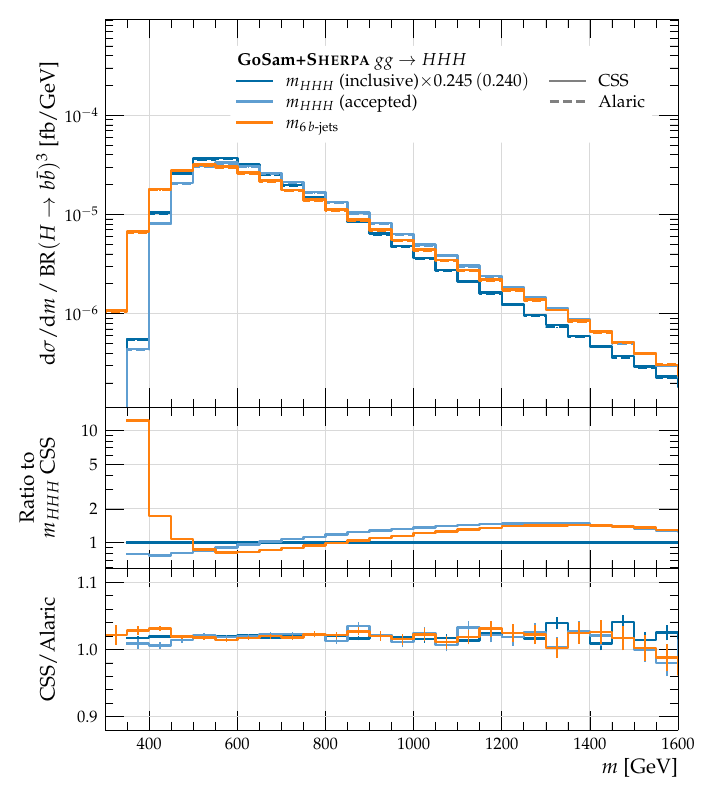}
    \includegraphics[width=0.49\linewidth]{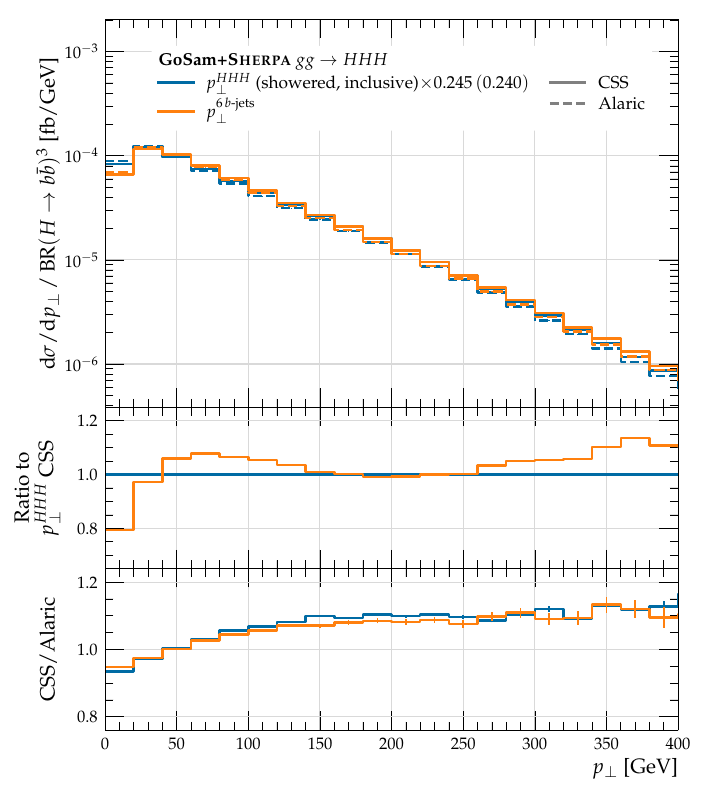}
    \caption{Shower effects on kinematical variables of the triple-Higgs system.
    Left: mass of the triple-Higgs system, measured inclusively with flat acceptance from \CSS~(\alaric), on events where the decay products pass the analysis cuts, and on the resulting 6 $b$-jets after the shower.
    Right: transverse momentum of the triple-Higgs system, measured inclusively with a flat acceptance from \CSS~(\alaric) and transverse momentum of the 6-b-jet system after the shower.}
    \label{fig:shower_mHHH}
\end{figure}

In Fig.~\ref{fig:shower_mHHH} we study kinematical variables of the triple-Higgs system.
Without the decays, the overall mass $m_{HHH}$ is unaffected by the parton shower.
We additionally plot the distribution measured on momenta of matrix-element level Higgs bosons in events that are accepted after decays and shower.
This reveals that the acceptance is somewhat lower than average around the threshold and then rises towards larger invariant masses.
The mass of the 6-b-jet system is affected by the same acceptance, but additionally by parton shower emissions outside the jets that lower their energies.
Consequently, the mass of this system can be significantly lower than the threshold for triple-Higgs production and is enhanced at small values.
At larger values, it tends towards the distribution measured on stable Higgs bosons.
The two showers agree to within a few per cent, with the difference being flat in $m_{HHH}$ and driven by the different acceptance rates.
On the right-hand side of the same figure, we show the transverse momentum of the same triple-Higgs and 6-b-jet systems.
At matrix-element level, at leading order, the transverse momentum vanishes trivially, so we only show the distribution after the parton shower, both with stable Higgs bosons and measured on the 6 $b$-jets.
We observe a larger dependence on the parton shower, amounting to a 10\% effect at large $p_\perp$, while the two distributions cross at lower values.
Large corrections can be expected since we do not include additional jets in the simulation, making this a prediction of the parton shower and its logarithmic approximation only.
Since the process under consideration is loop-induced at Born level, calculating this distribution even at its first non-trivial order would require two-loop amplitudes in the matching.
Alternatively, one could consider merging higher jet multiplicities.
Within the SM, this would be straightforwardly supported by the interface presented here.
However, we defer it and the discussion of HEFT contributions to future work.
The observations about the parton-shower dependence hold independently of whether the measurement is performed on stable Higgs bosons or $b$-jets.
The distribution itself is sensitive to this.
However, it is less straightforward to disentangle acceptance effects and genuine loss of radiation from the 6-b-jet system.
Overall, we observe an enhancement at large transverse momenta and a suppression in the first bin, relative to the acceptance-rate-scaled distribution measured on stable Higgs bosons.

\begin{figure}[!tb]
    \centering
    \includegraphics[width=0.49\linewidth]{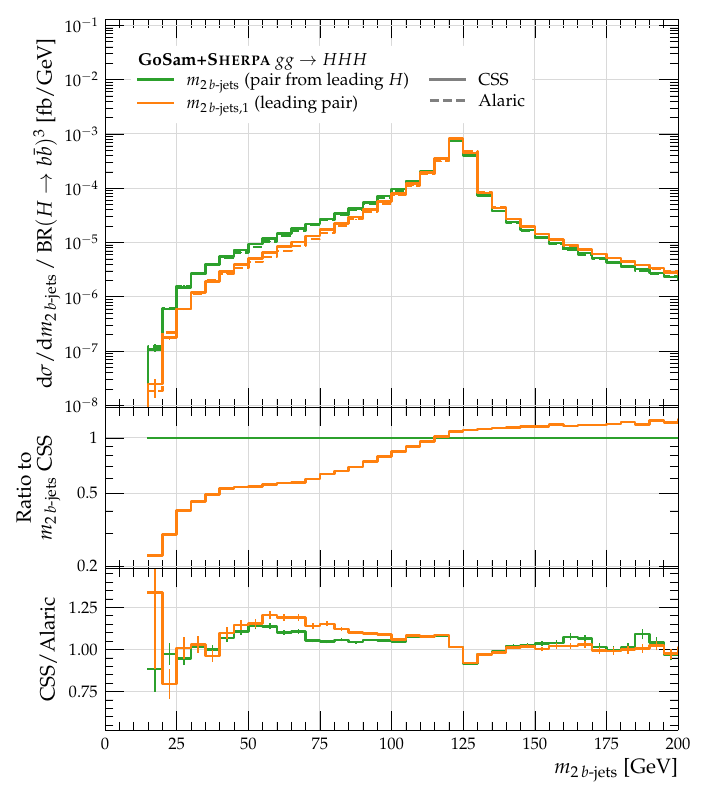}
    \includegraphics[width=0.49\linewidth]{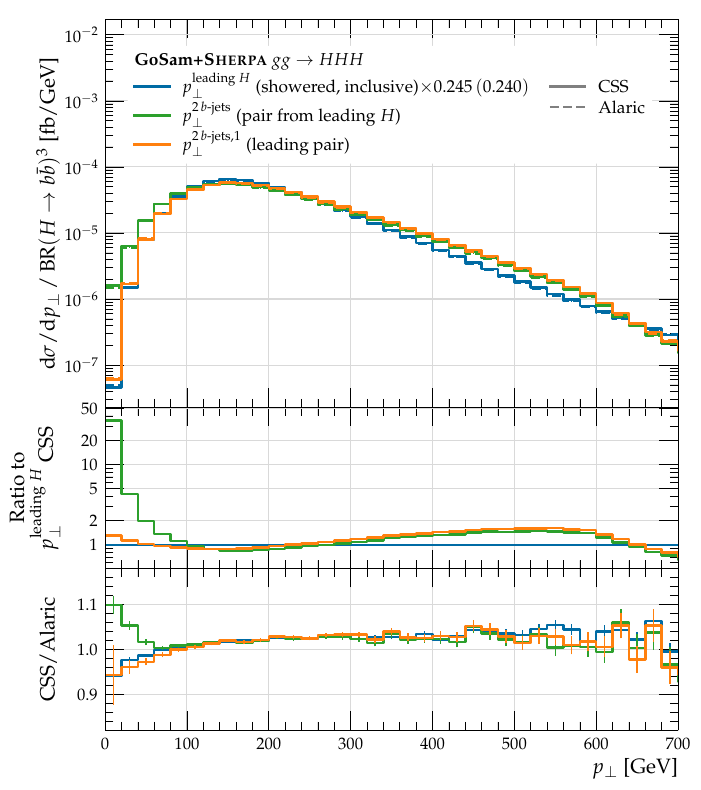}
    \caption{Shower effects on kinematical variables of the leading (in transverse momentum) Higgs boson and the resulting $b$-jets after the shower.
    Left: mass of the 2 $b$-jets resulting from the decay of the leading Higgs boson, and mass of the leading pair of $b$-jets matched to a Higgs boson.
    Right: transverse momentum of the leading Higgs, measured inclusively with a flat acceptance from \CSS~(\alaric), transverse momentum of the 2 $b$-jets resulting from the decay of the leading Higgs boson, and transverse momentum of the leading pair of $b$-jets matched to a Higgs boson.}
    \label{fig:shower_pTH}
\end{figure}

As another representative example, we consider the distributions of the Higgs boson with the largest transverse momentum in Fig.~\ref{fig:shower_pTH}.
In the left-hand plot, we show the invariant mass of the two corresponding $b$-jets.
The Higgs itself is generated on-shell, so neither it nor the unshowered decay products have a non-trivial distribution here.
We illustrate the mass of two selections of $b$-jets, making use of different amounts of Monte Carlo truth information.
First, we consider the pair of jets that contains the decay products of the Higgs boson with the largest transverse momentum at matrix-element level.
After the shower, these jets can potentially have lost part of their energy to radiation outside the jet radius, and other jets might have gained energy from initial-state radiation or clustering effects.
This means that the ordering in transverse momentum of the b-jet pairs is not necessarily the same, and we additionally show the mass distribution of the pair with the largest transverse momentum after the shower.
Note, however, that we still only consider pairs of $b$-jets originating from the same Higgs decay using Monte Carlo truth information, so the change is purely a measure of the energy lost due to radiation.
In the right-hand plot of Fig.~\ref{fig:shower_pTH} we show the transverse momentum of the leading Higgs boson, again measured from either the 2 $b$-jets from the leading matrix-element Higgs or those that form the leading tagged b-jet pair.
In this case, we additionally show the distribution of the transverse momentum of the leading Higgs boson in the sample without decays, scaled by the acceptance factors.
In both observables, small values potentially imply a large loss of energy to radiation outside the jets.
Consequently, the distributions measured on the decay products of the leading matrix-element Higgs boson are strongly enhanced there, since jets here can evolve to small values without being discarded.
On the other hand, when considering the leading b-jet pair, a large energy loss will typically imply that a different pair of jets becomes the leading one, so small values are suppressed.
The transverse momentum distribution as well as the mass distribution at masses above the Higgs mass show only small variations between the showers, of the order of a few per cent.
Below the Higgs mass, the spectrum is affected more strongly, reaching up to 10-20\% depending on the jet selection.
\begin{figure}[!htb]
    \centering
    \includegraphics[width=0.49\linewidth]{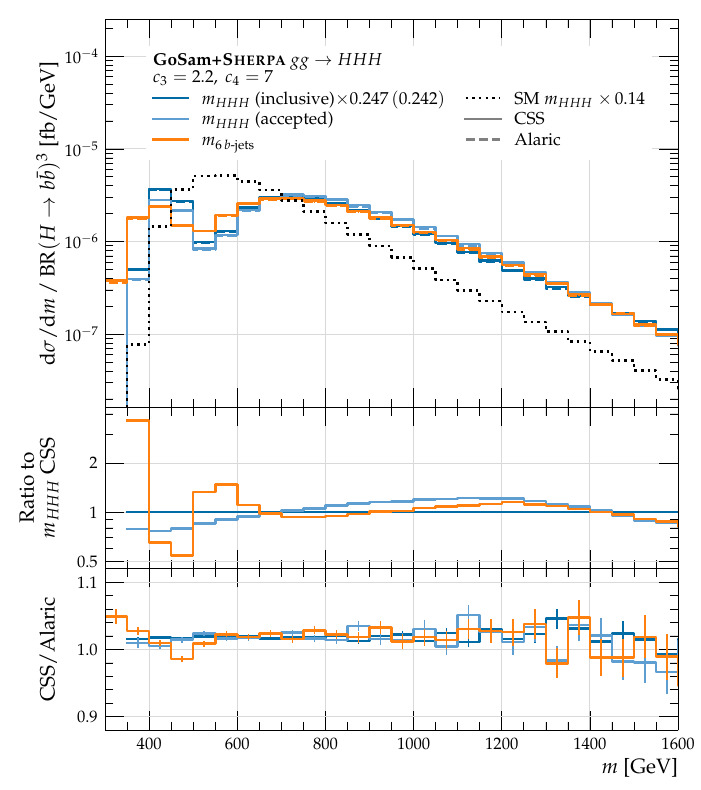}
    \includegraphics[width=0.49\linewidth]{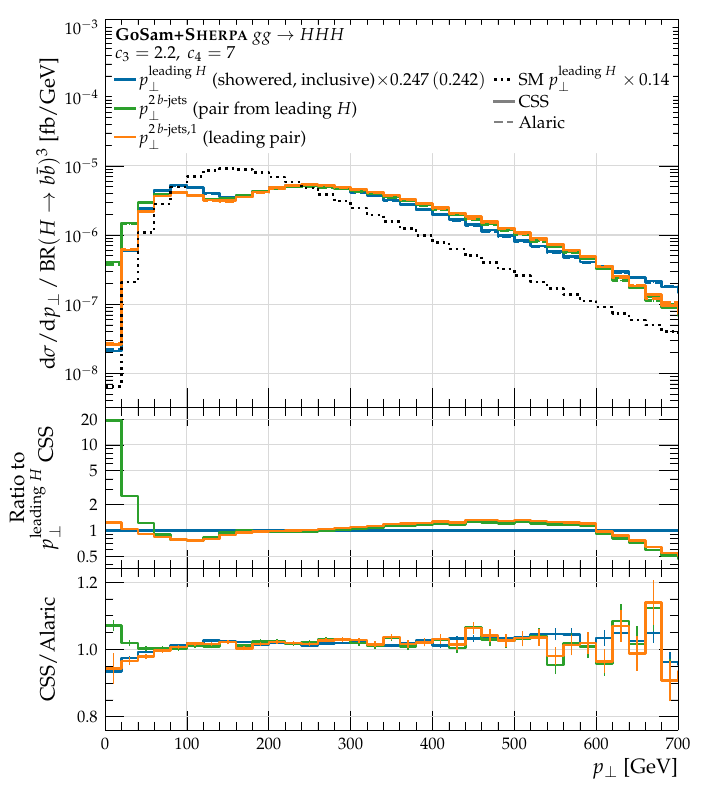}
    \includegraphics[width=0.49\linewidth]{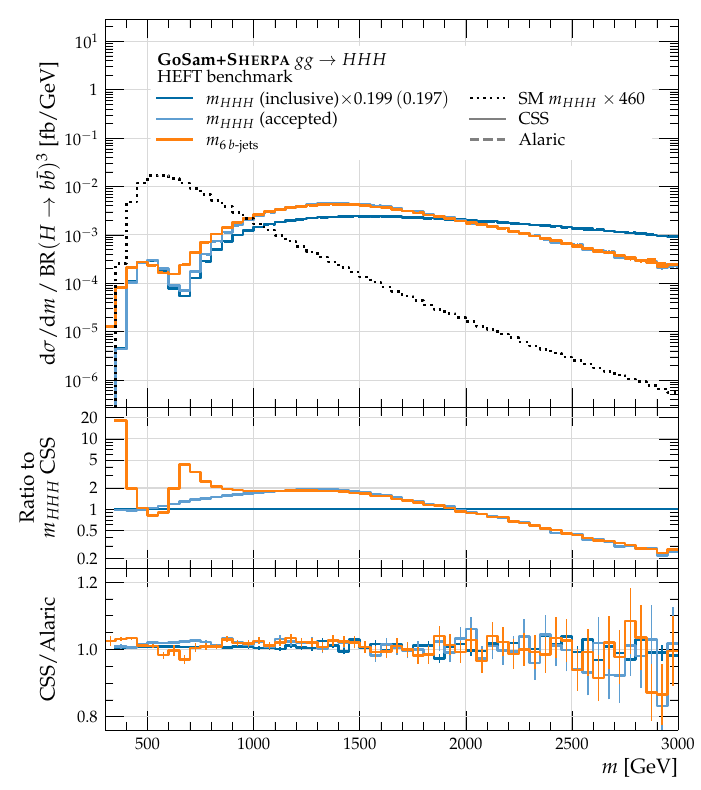}
    \includegraphics[width=0.49\linewidth]{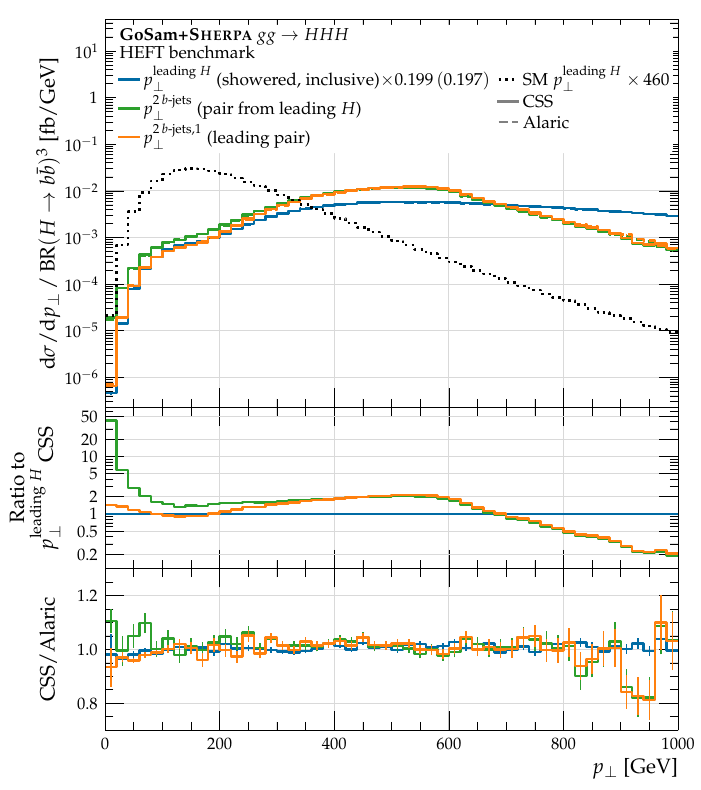}
    \caption{Shower effects on the mass of the triple-Higgs system and the transverse momentum of the leading Higgs boson, compared to the effect of non-zero HEFT couplings.
    Left: mass of the triple-Higgs system, measured inclusively with flat acceptance from \CSS~(\alaric), on events where the decay products pass the analysis cuts, and on the resulting 6 $b$-jets after the shower.
    Right: transverse momentum of the leading Higgs boson, measured inclusively with a flat acceptance from \CSS~(\alaric) and transverse momentum of the 6-b-jet system after the shower.}
    \label{fig:shower_vs_heft}
\end{figure}

We finally compare the effects of the parton shower to the distortions of the triple-Higgs mass and the leading Higgs boson transverse momentum spectrum.
In Fig.~\ref{fig:shower_vs_heft} we show plots for both observables, in the same shower configurations as before, for the benchmark points investigated in the previous section.
The inclusive curves are again multiplied by a shower-dependent acceptance factor, that in the HEFT scenarios slightly changes.
We also show the SM result for stable Higgs bosons as a reference, scaled to match the cross section of the respective HEFT scenario.
In both cases, the dip in the $m_{HHH}$ spectrum is washed out when measured on $b$-jets, while it is not affected by the shape of the acceptance for our cuts.
At larger masses, the spectrum is suppressed relative to the stable result by the acceptance.
This effect is also visible in the SM plots, however there the spectrum is concentrated at smaller values and we terminated the plot before the effect grew too large.
We interpret this as the approach of the boosted regime, where it is increasingly unlikely that all $b$-quarks are in separate jets as we require in our analysis.
A phenomenological analysis of this regime would need a dedicated analysis where fat jets would be tagged as Higgs jets, or even multi-Higgs ones, based on their constituents.
We do not consider this possibility here.
All conclusions for the HEFT points hold without any significant modifications from changing the parton shower.
\FloatBarrier

\section{Conclusions}
\label{sec:conclusions}
We have presented a calculation of the loop-induced $gg\to HHH$ process in the Higgs Effective Field Theory (HEFT), including the effect of parton showers off the initial state and off the Higgs decay products.
To achieve this, we combined the \gosam matrix element generator with the general-purpose event generator \sherpa, based on a HEFT model file in UFO format.

We studied the dependence of the $gg\to HHH$ cross section at the LHC as a function of the introduced HEFT couplings.
We found that both total cross sections and differential observables such as the triple Higgs invariant mass distribution are quite sensitive to changes of the HEFT couplings, in particular also to the effective multi-Higgs-gluon couplings that have been neglected in the literature so far for this process.

On the differential level, we studied representative kinematic observables of the triple-Higgs system and of the Higgs boson with the largest transverse momentum.
We found that both can be reshaped non-trivially by HEFT operators.
These features are washed out to an extent but survive the effect of parton showers in general.
More phenomenological studies are possible based on this framework, and should be carried out with the requirements of the LHC experiments in mind, especially regarding tagging of $b$-jets, practical assignment of jets (or decay products in general) to Higgs bosons, and investigation of the boosted regime with jet substructure techniques.

Finally, we stress that the \textsc{GoSam}+\sherpa setup we have developed here is not restricted to this particular process or to HEFT.
Any model supported in \gosam, in particular models in UFO format, could be used at leading order, potentially including loop-induced processes in these models, as we did in this work.
Likewise, one could study other loop-induced processes in the SM through the interface in much the same way.
Especially, this opens an effective way to produce results for loop-induced processes with additional jets that could be used in merging for such processes.
We anticipate it to be used in a number of studies that were not possible in an automated way before.
Further, it is possible to use \gosam as a one-loop provider in \sherpa for Standard Model processes.
In the future, we plan to extend the interface to include the possibility of automated NLO calculations for non-SM processes.

\acknowledgments

All Feynman diagrams in this note are drawn using \texttt{FeynGame} \cite{Bundgen:2025utt}.

We would like to thank Charlotte Kern and Augustin Vestner for useful discussions.
%We also thank Michelle Gensmann and Ulrich Haisch for useful exchange related to the comparison of results. 
We also thank Margarete M\"uhlleitner for providing the version of \textsc{Hpair} which we used for validation of the CP-violating couplings.
J.B. is supported in part by the BMFTR under grant agreement  05H24VKB. G.H. is supported in part by the German Research Foundation (DFG) under grant 396021762 - TRR 257. 
D.R. is supported by the European Union under the HORIZON program in Marie Sk{\l}odowska-Curie project No. 101153541.

\appendix

\section{Parameterisation of the Total Cross Section}
\label{sec:app:parameterisation}

Each appearance of a coupling in a Feynman diagram introduces one factor of this coupling in the expression of the respective diagram. Therefore, each diagram is a monomial in a specific combinations in the coupling, and the matrix element a polynomial in the couplings. With the 84 diagrams contributing to this process, eleven combinations appear in the matrix element,
\begin{equation}
\begin{aligned}
    \mathcal{M} =& a_0 c_{3g} + a_1 c_t c_{2t} + a_2 c_3^2 c_t + a_3 c_3 c_t^2 + a_4 c_3 c_{2t} \\
    &+ a_5 c_{3t} + a_6 c_t^3 + a_7 c_4 c_g + a_8 c_t c_4 + a_9 c_3 c_{2g} + a_{10} c_3^2 c_g.
\end{aligned}
\end{equation}
After squaring the matrix element, 62 unique combinations remain, resulting in the following parameterisation of the total cross section:
\begin{equation} \label{eq:xsec}
\begin{aligned}
   \frac{\sigma_\mathrm{HEFT}}{\sigma_\mathrm{SM}} = 
   & A_{0} c_3 c_{2g} c_t^{3}+A_{1} c_{3g} c_t^{3}+A_{2} c_3^{2} c_g c_{3t}+A_{3} c_4 c_g c_{3t}+A_{4} c_3^{2} c_{2g}^{2}+A_{5} c_3^{2} c_g c_t c_{2t}+A_{6} c_4 c_t^{4}\\
   &+A_{7} c_4^{2} c_t^{2}+A_{8} c_3 c_4 c_g c_t^{2}+A_{9} c_3^{2} c_4 c_g c_t+A_{10} c_3^{2} c_t c_{3t}+A_{11} c_4 c_t c_{3t}+A_{12} c_3^{2} c_{2g} c_t^{2}\\
   &+A_{13} c_3^{4} c_g c_t+A_{14} c_3 c_t^{5}+A_{15} c_3 c_4 c_g c_{2t}+A_{16} c_3 c_{2g} c_t c_{2t}+A_{17} c_3^{2} c_{2g} c_{2t}+A_{18} c_3^{3} c_g c_t^{2}\\
   &+A_{19} c_4 c_g c_t c_{2t}+A_{20} c_4 c_g c_{3g}+A_{21} c_3^{2} c_g c_{3g}+A_{22} c_3 c_{2g} c_{3t}+A_{23} c_{3t}^{2}+A_{24} c_{3g} c_{3t}\\
   &+A_{25} c_3 c_4 c_g c_{2g}+A_{26} c_3 c_4 c_t c_{2t}+A_{27} c_3^{3} c_g c_{2t}+A_{28} c_3^{2} c_4 c_g^{2}+A_{29} c_3^{4} c_g^{2}+A_{30} c_4^{2} c_g c_t\\
   &+A_{31} c_3 c_4 c_t^{3}+A_{32} c_3^{3} c_t c_{2t}+A_{33} c_3^{2} c_g c_t^{3}+A_{34} c_4 c_g c_t^{3}+A_{35} c_t^{4} c_{2t}+A_{36} c_3 c_t c_{2t}^{2}\\
   &+A_{37} c_t c_{2t} c_{3t}+A_{38} c_t^{6}+A_{39} c_3 c_{3g} c_t^{2}+A_{40} c_t^{3} c_{3t}+A_{41} c_3^{3} c_g c_{2g}+A_{42} c_4^{2} c_g^{2}\\
   &+A_{43} c_{3g} c_t c_{2t}+A_{44} c_3^{3} c_t^{3}+A_{45} c_3^{2} c_t^{4}+A_{46} c_3 c_{2g} c_{3g}+A_{47} c_3 c_{3g} c_{2t}+A_{48} c_3^{2} c_{2t}^{2}+A_{49} c_{3g}^{2}\\
   &+A_{50} c_3 c_4 c_{2g} c_t+A_{51} c_3 c_{2t} c_{3t}+A_{52} c_3^{2} c_{3g} c_t+A_{53} c_4 c_{3g} c_t+A_{54} c_3^{2} c_4 c_t^{2}+A_{55} c_3 c_t^{3} c_{2t}\\
   &+A_{56} c_t^{2} c_{2t}^{2}+A_{57} c_3^{2} c_t^{2} c_{2t}+A_{58} c_4 c_t^{2} c_{2t}+A_{59} c_3^{3} c_{2g} c_t+A_{60} c_3^{4} c_t^{2}+A_{61} c_3 c_t^{2} c_{3t}.
\end{aligned}
\end{equation}

% Bibliography

\bibliographystyle{JHEP}
\bibliography{biblio.bib}

\end{document}